\RequirePackage[l2tabu, orthodox]{nag}
\RequirePackage{snapshot}

\documentclass[9pt,onecolumn]{extarticle}

\makeatletter
\if@twocolumn
  \usepackage[dvips,letterpaper,top=0.5in, bottom=0.5in, left=0.75in, right=0.5in,includefoot,heightrounded]{geometry}
\else
  \usepackage[dvips,letterpaper,margin=1in,includefoot,heightrounded]{geometry}
\fi

\usepackage{srcltx}

\usepackage[russian,portuges,english]{babel}

\iflanguage{portuges}
    {\newcommand{\keywordname}{Palavras-chaves}}
    {\newcommand{\keywordname}{Keywords}}

\usepackage{amsmath}
\usepackage{amssymb,amsfonts}

\usepackage{abstract}

\usepackage{graphicx}
\usepackage{epsfig}
\usepackage[usenames,dvipsnames,svgnames,x11names]{xcolor}
\usepackage{subfigure}

\usepackage{booktabs}

\usepackage{setspace}
\usepackage{flushend}

\usepackage{cite}

\usepackage{hyperref}
\usepackage[normalem]{ulem}

\usepackage{enumerate}

\usepackage{multirow}
\usepackage[noend]{algpseudocode}

\usepackage{listings}

\usepackage[short,12hr]{datetime}

\usepackage{psfrag}

       \usepackage{fouriernc}
\makeatletter

\newenvironment{rsmallmatrix}{\null\,\vcenter\bgroup
  \Let@\restore@math@cr\default@tag
  \baselineskip6\ex@ \lineskip1.5\ex@ \lineskiplimit\lineskip
  \ialign\bgroup\hfil$\m@th\scriptstyle##$&&\thickspace\hfil
  $\m@th\scriptstyle##$\crcr
}{%
  \crcr\egroup\egroup\,%
}

\newcommand{\printtitle}{%
\makeatletter
\if@twocolumn

\twocolumn[%
  \maketitle
  \begin{onecolabstract}
    \myabstract
  \end{onecolabstract}
  \begin{center}
    \small
    \textbf{\keywordname}
    \\\medskip
    \mykeywords
  \end{center}
  \bigskip
]
\saythanks
\else
  \maketitle
  \begin{onecolabstract}
    \myabstract
  \end{onecolabstract}
  \begin{center}
    \small
    \textbf{\keywordname}
    \\\medskip
    \mykeywords
  \end{center}
  \bigskip
  \onehalfspacing
\fi
\makeatother
}

\author{%
A. Madanayake%
\thanks{
Electrical and Computer Engineering Department, Florida International University, USA (e-mail:~\url{amadanay@fiu.edu},~\url{akram.m.n@ieee.org},~\url{pberu002@fiu.edu})
}
\and
R.~J.~Cintra%
\thanks{%
Signal Processing Group, UFPE, Brazil;
and
School of Science and Mathematics, Howard Payne University, TX, USA. (e-mail: \url{rjdsc@de.ufpe.br})}
\and
N. Akram$^\ast$
\and
V. Ariyarathna$^\ast$
\and
S. Mandal%
\thanks{Electrical, Computer, and Systems Engineering Department, Case Western Reserve University, USA (e-mail:~\url{sxm833@case.edu})}
\and
V.~A.~Coutinho%
\thanks{Universidade Federal Rural de Pernambuco, Recife, Brazil (e-mail: \url{vitor.coutinho@ufrpe.br})}
\and
F.~M.~Bayer%
\thanks{Department of Statistics
and LACESM,
Federal University of Santa Maria, Brazil (e-mail:~\url{bayer@ufsm.br})}
\and
D. Coelho%
\thanks{Independent researcher, Calgary, Alberta, Canada (e-mail: \url{diegofgcoelho@gmail.com})}
\and
T. S. Rappaport%
\thanks{NYU Wireless, Department of Electrical Engineering, Tandon School of Engineering, New York University, Brooklyn NY 11220, USA (e-mail: \url{tsr@nyu.edu})}

}

\title{%
Fast Radix-32 Approximate DFTs for 1024-Beam Digital RF Beamforming}

\newcommand{\myabstract}{%
The discrete Fourier transform (DFT)
is widely employed for multi-beam digital
beamforming.
The DFT can be efficiently implemented through
the use of fast Fourier transform~(FFT)
algorithms, thus
reducing chip area, power consumption, processing time,
and consumption of other hardware resources.
This paper proposes three new hybrid DFT 1024-point DFT approximations
and their respective fast algorithms.
These approximate DFT (ADFT) algorithms have
significantly reduced circuit complexity and power consumption
compared to traditional FFT
approaches while trading off a subtle loss in computational precision
which is acceptable for digital beamforming applications in RF antenna implementations. ADFT algorithms have not been introduced for beamforming beyond $N =  32$, but this paper anticipates the need for massively large adaptive arrays for future 5G and 6G systems.
Digital CMOS circuit designs for the ADFTs
show the resulting improvements in both circuit complexity
and power consumption metrics.
Simulation results
show similar or lower critical path delay
with up to 48.5\% lower chip area compared
to a standard Cooley-Tukey FFT.
The time-area and dynamic power metrics are
 reduced up to 66.0\%. The 1024-point ADFT beamformers produce signal-to-noise ratio (SNR) gains between~29.2--30.1 dB, which is a loss of~$\le$~0.9 dB SNR gain compared to exact 1024-point DFT beamformers (worst case) realizable at using an FFT.
}

\newcommand{\mykeywords}{%
Fourier Transform, Discrete Fourier Transform, Approximation, VLSI, DFT, FFT, Radix, Fast Algorithm, Beamforming, Beamsteering, 5G, MIMO, Massive MIMO.
}

\date{}

\begin{document}

\printtitle

{T}{he} discrete Fourier transform~(DFT) is a linear transform
that is widely applied to convert a sampled signal into a representation
over the discrete frequency domain. Fully-digital transmit and receive aperture arrays for radio-frequency (RF) spectrum sensing,
communications,
and radar use the DFT for multi-beam beamforming.
For example, simultaneous
receiver beams are imperative for high-capacity multi-input multi-output~(MIMO) wireless communication systems.
Joint spatial division and multiplexing~(JSDM) is an approach to multi-user MIMO downlinks that exploits
the structure of  channel correlations in order to allow a large number of antennas at the base station while
requiring reduced-dimensional channel state information at the
transmitter~\cite{zhang2015large, shu2014}.
This uses a multi-user MIMO downlink precoder obtained from an array pre-beamforming matrix, and incurs no loss of
optimality for a large number of array  elements. A DFT-based
pre-beamforming matrix is near-optimal for uniform linear arrays (ULAs) of antennas, and requires only coarse
information about the users' angles of arrival and angular spread~\cite{adhikary2013}.

\begin{figure*}[t!]
\centering
\scalebox{1.5}{\includegraphics{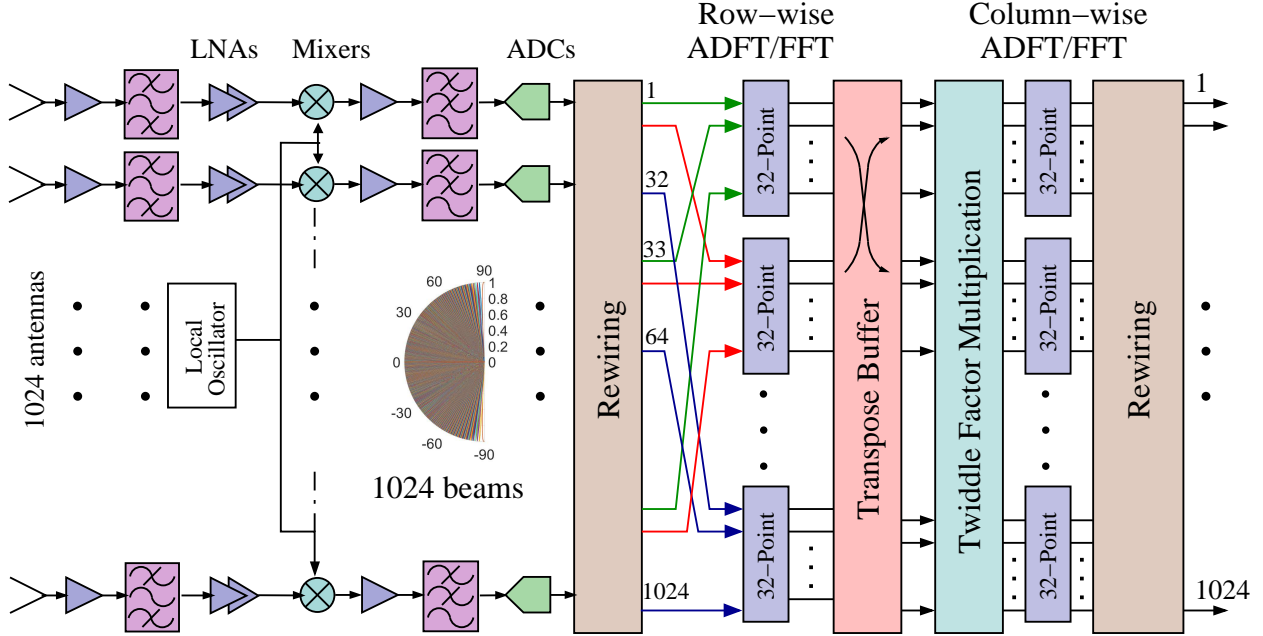}}
\caption{Beamforming architecture of a 1024-element ULA receiver using the proposed method. The rewiring block performs spatial multiplexing over the incoming in the input and the transformed signal at the output (see Fig.~\ref{algorithm_block}).\label{fig:beamforming_sch} }
\end{figure*}
The $N$-point DFT computes
$N$~uniformly spaced frequency domain outputs~(``bins'')
using $N$~uniformly sampled discrete signal values
by means of
an $N\times N$
transform matrix~\cite{champeney1987handbook}.
Because implementations of the multiplication operation requires  more chip area (or processing time, for non-parallelized software implementations) compared to addition operations,
the computational complexity of computing the DFT is
expressed in terms of the  multiplication count~\cite{blahut2010fast}.
The required number of multiplications
depends on the fast algorithm
employed for the particular
transform length~$N$ in consideration.
The computational complexity of the $N$-point DFT
using direct matrix-vector multiplication is~$\mathcal{O}(N^2)$
where~$\mathcal{O}(\cdot)$
represents the ``big O'' notation for asymptotic complexity~\cite{Oppenheim2009, Madanayake2018, Ariyarathna2019}.

The
computational complexity of computing the $N$-point DFT
can be reduced via fast Fourier transforms~(FFTs), which are fast algorithms for realizing DFTs that reduce the computational complexity to~$\mathcal{O}(N\log_2 N)$~\cite{blahut2010fast}. Thus, multiple DFT beams for both wireless communications applications (e.g., JSDM) and multi-beam radar/imaging systems are  often generated by applying an $N$-point spatial FFT to each temporal sample acquired by
the ULA~\cite{Ellingson2003,coleman2007generalized}.

The search for particular $N$-point FFT methods
that minimize the multiplicative
complexity
is a separate field of research in signal processing,
computer science, and applied mathematics,
with a multitude of algorithms and implementations
available~\cite{blahut2010fast, ref1, ref2, ref3, ref4}.
In~\cite{heideman},
the theoretical lower bound
for the DFT multiplicative
complexity was established as a function of~$N$.
All FFT algorithms use sparse factorizations
of the DFT matrix to provide accurate implementations of the DFT at an arithmetic complexity that approaches this lower bound.
However, such high accuracy is of limited practical relevance
in digital multi-beam RF beamforming applications, such as radar signal processing,
where the accuracy of the results is limited by other system parameters or environmental conditions
(e.g., thermal noise in a receiver, or the practical implementation of an antenna radiation pattern compared to ideal, harmonic distortion
in a microwave mixer or amplifier).
In such applications,
relentless pursuit of high accuracy in the
exact computation of the DFT is not relevant in terms of
overall performance, and smart system designers can exploit this fact for power and cost optimization.
High-precision VLSI implementation of FFT algorithms
may result in unnecessarily large circuits,
exaggerated critical path delays,
and wasted power.
All of those factors contribute to higher-cost circuits,
reduced frequency of operation,
and higher operation costs.
This is because digital multipliers
demand a large amount of circuit
resources when compared to simple adders.
This makes the reduction of the number of multipliers in
a given system crucial
when chip area and power must be
conserved and high-speed operation is desirable.
In particular, the
adoption of approximate DFT (ADFT)
computations opens up new possibilities
for fast algorithms which
do not compute the DFT
in the strictest mathematical sense,
but nevertheless can be \emph{good enough}
for digital multi-beam RF beamforming applications, particularly at mmwave frequencies and above, where reproducibility of antenna patterns become more problematic.
Because ADFT applications are able to realize much greater efficiencies than the theoretical lower bound $N$ for an $N$-point DFT proposed in~\cite{heideman},
ADFT computations allow greater reductions in computational complexity than traditional FFTs,
albeit at the cost of a deterministic loss in performance, namely a small increase in worst-case side-lobe level~\cite{ariyarathna2018analog}.

The ever increasing data rate demands of wireless communications led to the exploration of millimeter-wave (mmW)/sub-THz/THz frequencies in 5G cellular networks~\cite{rappaport2013,rap2019}, where larger antenna array sizes (e.g. $N =64,~ 128,~256$) for beamforming and massive MIMO have become a general requirement \cite{BJORNSON20193}.
For example, IoT and robotics applications in emerging fifth-generation (5G) and beyond mobile wireless networks will require 6D positioning, which involves both spatial position and device orientation (role, pitch, yaw) which require new algorithms that can benefit from large number of closely packed low-complexity digital beams \cite{6D_1,6D_2,shu2014}.
A similar need occurs in the design of
systems for intelligent surfaces which provide means of communication
without line-of-sight~\cite{taha2019enabling}.
In fact mmW-based 5G MIMO cellular systems are already being deployed~\cite{5Gnews}.
Moreover, ongoing research in the sub-THz range~\cite{wb1,wb2,wb3,wb4,wb5,wb6,wb7,wb8,wb9,shu2014} suggests that
the W and G bands will be commercially available within the next 5-10 years.
Such sub-THz carrier frequencies require large amounts of beamforming gain to mitigate free-space path loss in the first meter of propagation from the antenna~\cite{rap2019,shu2014,Globecom2018rap,Globecom2019rap}. Thus, communication systems at these frequencies would require much larger  numbers of antenna elements in the transceiver arrays; array sizes of the order of $N=1000$ elements would not be unrealistic for future sixth generation~(6G) cellular systems.
Nevertheless,
to the best of our knowledge,
DFT approximations in the literature are limited to $N\le32$.

In this paper,
we address this important beamforming challenge by introducing three
new approximations to the very large N = 1024 (1024-point) DFT.
Fast algorithms that allow
low-complexity implementations of these approximations are also developed and shown to provide remarkable accuracy with significant cost and power reduction compared to DFT and FFT approaches.
The proposed 1024-point ADFTs
are based on a recently proposed 32-point DFT approximation
and multiplierless fast algorithm~\cite{DoraThesis,Madanayake2018}
that furnish a ``reasonable'' approximation of the 32-point DFT
albeit without using multiplications
(i.e., using an adder-only signal flow graph).
The 1024-point exact DFT can be expressed in terms of 32-point DFT. We use this fact to derive an approximation for the 1024-point DFT matrix by means for our earlier 32-point ADFT.
In particular, we propose three different 1024-point transforms with different trade-offs in computational complexity and
computational accuracy compared to the baseline exact DFT. These three transforms differ from each other based on the use of 32-point ADFT in the derivation and they can be used to replace the FFT while generating $N=1024$ beams from a 1024-element ULA as shown in Fig.~\ref{fig:beamforming_sch}.

The paper is organized as follows.
Section~\ref{s:review} reviews the DFT and selected popular FFT algorithms.
In Section~\ref{sec:32point},
we discuss the mathematical background
for the 32-point DFT approximation
introduced in~\cite{DoraThesis}
and
describe its associated fast algorithm
in matrix form.
In Section~\ref{s:1024},
we present 1024-point DFT approximations
and
discuss three different algorithms to implement them.
Section~\ref{s:vlsi}
explores the digital VLSI realization
of the proposed 1024-point DFT approximations.
In Section~\ref{s:conclusion},
we summarize our conclusions.

\section{Review of the DFT and FFT}\label{s:review}
In order to understand the
method used to create
accurate ADFT algorithms,  we will discuss the mathematical background
related to the DFT definition and FFT algorithms.
\subsection{Mathematical Definition of the DFT} \label{sub:dft}

Let the vector~$\mathbf{x} = \begin{bmatrix} x[0] & x[1] & \ldots & x[N-1] \end{bmatrix}^\top$
represent a signal with~$N$ samples.
The DFT
maps the input signal $\mathbf{x}$
into an output signal
$\mathbf{X} =
\begin{bmatrix}
X[0] & X[1] & \cdots & X[N-1]
\end{bmatrix}^\top$
according to the following relationship:
\begin{align}
X[k] &
\triangleq
\frac{1}{\sqrt{N}}
\sum_{n=0}^{N-1}
x[n]
\cdot
\omega_N^{nk}
,
\quad
k = 0, 1, \ldots, N-1,
\end{align}
where
$\omega_N = e^{-j\frac{2\pi }{N}}$
is the $N$th root of unity
and $j\triangleq\sqrt{-1}$.
On the other hand,
the inverse DFT (IDFT) is
given as
\begin{align}
x[n] &= \frac{1}{\sqrt{N}}
\sum_{k=0}^{N-1}
X[k]
\cdot
\omega_N^{-nk}
,
\quad
n = 0, 1, \ldots, N-1.
\end{align}
The DFT of~$\mathbf{x}$
can be expressed through a matrix-vector multiplication
$\mathbf{X} = \mathbf{F}_N \cdot \mathbf{x}$,
where
\begin{align}
\mathbf{F}_N
=\frac{1}{\sqrt{N}}\left[
  \begin{smallmatrix}
    1 & 1 & 1 &  \ldots  & 1  \\
    1 & \omega_N & \omega_N^2 &  \ldots  &  \omega_N^{(N-1)}  \\
    1 & \omega_N^2 & \omega_N^4 &  \ldots  & \omega_N^{2(N-1)}  \\
    1 & \omega_N^3 & \omega_N^6 &  \ldots  & \omega_N^{3(N-1)}  \\
    \vdots & \vdots & \vdots &   \ddots & \vdots\\
    1 & \omega_N^{(N-1)} & \omega_N^{2(N-1)}&  \ldots  & \omega_N^{(N-1)(N-1)}  \\
  \end{smallmatrix}\right]
\end{align}
is the $N$-point DFT matrix~\cite{rao}.

\subsection{FFT Algorithms}
DFT was originally the cornerstone of primitive DSP, until the FFT was found to be vastly more efficient. Here we extend FFTs to become ADFTs. The computational complexity associated with performing
the $N$-point DFT
operation in direct form is~$\mathcal{O}(N^2)$.
This complexity is prohibitive for most engineering applications
since
a high number of operations accounts for
(i)~higher energy consumption;
(ii)~higher latency;
(iii)~higher number of gates;
and, in consequence,
(iv)~higher chance of system failure.
To address these issues,
FFT factorizations
furnish
a product of sparse (mostly zeros) matrices
that
reduces
the DFT computational complexity to~$\mathcal{O}(N\log N)$.
Different FFT algorithms
can be identified
in the literature~\cite{CoTu,Duhamel1984,W1,W2}.
Here we consider three popular algorithms, namely
i) the Cooley-Tukey FFT~\cite{blahut2010fast},
ii) the split-radix FFT~\cite{Duhamel1984},
and
iii) the Winograd FFT~\cite{Winograd1980};
each of these is briefly described below.

\subsubsection{Cooley-Tukey Algorithm}

A very popular form of the classical Cooley-Tukey algorithm
is the radix-2 decimation-in-time FFT,
which splits the $N$-point DFT computation into
two $N/2$-point DFT computations
resulting in an overall reduced complexity~\cite{CoTu}.
Recursive use of this algorithm
reduces the number of
multiplications of the DFT from $\mathcal{O}(N^2)$ down
to $\mathcal{O}(N\log_2N)$.

\subsubsection{Split-radix Algorithm}
This is a variant of the Cooley-Tukey
FFT algorithm which
uses a blend of radix-2 and radix-4 by
recursively expressing the
$N$-point DFT
in terms of
one $N/2$-point DFT
and
two $N/4$-point DFT instantiations~\cite{Duhamel1984}.
The split-radix algorithm
can reduce the overall number
of additions required to compute DFTs
of sizes that are powers of two without increasing
the number of multiplications~\cite{johnson2007modified}.

\subsubsection{Winograd Algorithm}
The Winograd algorithm implements an efficient FFT and exploits
the multiplicative structure on
the data indexing of DFT and converts it into a
cyclic convolution computation~\cite{W1,W2}.
In several particular cases,
the Winograd algorithm
achieves the theoretical minimum
multiplicative complexity~\cite{heideman}
as shown in~\cite{W1} making it more efficient over the Cooley-Tukey and radix.
For large DFT block lengths
that can be decomposed
as a product of small primes,
the Winograd algorithm achieves nearly-linear complexity~\cite{blahut2010fast}.

\subsection{Matrix Representation of
the $N^2$-point DFT in terms of
the $N$-point DFT}
\label{sec:DFT-N-squared}

Now we will use the matrix definition in sub section \ref{sub:dft} to derive a matrix representation for the computation of the
$N^2$-point DFT in terms
of the $N$-point DFT
via a radix-$N$ FFT approach. The goal of this is to derive a 1024-point DFT in terms of 32-point DFT.
Generally speaking,
the $N^2$-point DFT computation
corresponds to a vector-matrix multiplication
with a $N^2\times N^2$ matrix transformation:
\begin{align}
\label{equation-dft-def-n-squared}
\mathbf{X}
=
\mathbf{F}_{N^2}
\cdot
\mathbf{x}
.
\end{align}
The expression in~\eqref{equation-dft-def-n-squared}
can be rewritten by
directly invoking
the Cooley-Tukey algorithm
in its more general form
as detailed in~\cite[p.~69]{blahut2010fast}.
By explicitly following the Cooley-Tukey algorithm,
the $N^2$-point DFT
can be computed by means
of:
\begin{enumerate}
\item
address-shuffling the input column vector into a 2D $N\times N$ array;
\item
computing the $N$-point DFT of each array column using FFTs;
\item
element-wise multiplying the twiddle-factors (twiddle factors are the coefficients containing roots of unity in the DFT matrix~\cite{blahut2010fast});
\item
computing the $N$-point DFT of each resulting row using FFTs;
and
\item
undoing the address shuffling to convert the obtained 2D array
into the final output column vector.
\end{enumerate}
The 1D to 2D mapping can
be accomplished by means of the
inverse vectorization operator
$\operatorname{invvec}(\cdot)$~\cite{duong2019vec}
(Cf.~\cite{mathworks2019matlab,mathworks2019matlaba})
which
obeys the following mapping:
\begin{align}
\label{eq:mapping}
\operatorname{invvec}
\left(
\begin{bmatrix}
x_0 \\ x_1 \\ \vdots \\ x_{N^2}
\end{bmatrix}
\right)
=
\begin{bmatrix}
x_0     & x_N      & \cdots  & x_{N(N-1)}    \\
x_1     & x_{N+1}  & \cdots  & x_{N(N-1)+1}  \\
\vdots  & \vdots   & \ddots  & \vdots        \\
x_{N-1} & x_{2N-1} & \cdots  & x_{N^2-1}
\end{bmatrix}
.
\end{align}
Based on the 1D to 2D mapping in Eqn. \eqref{eq:mapping} we can show that the $N^2$-point DFT
given in~\eqref{equation-dft-def-n-squared}
can be represented in the following matrix expression
based on the Cooley-Tukey algorithm:
\begin{align}
\label{equation-ct-direct}
\mathbf{X}
=
\operatorname{vec}
\left(
\left\{
\mathbf{F}_N
\cdot
\left[
\mathbf{\Omega}_N
\circ
\left(
\mathbf{F}_N
\cdot
(\operatorname{invvec}(\mathbf{x}))^\top
\right)
\right]^\top
\right\}^\top
\right)
,
\end{align}
where
$\operatorname{vec}(\cdot)$
is the matrix vectorization operator~\cite[p.~239]{seber2008matrix},
$\circ$
is the Hadamard element-wise multiplication~\cite[p.~251]{seber2008matrix},
the superscript~$^\top$ denotes simple transposition (non Hermitian),
and
$\mathbf{\Omega}_N$ is the twiddle-factor matrix
given by
$\mathbf{\Omega}_N = (\omega_{N^2}^{m\cdot n})_{m,n=0,1,\ldots,N}$.
Noting that~$\mathbf{\Omega}_N^\top = \mathbf{\Omega}_N$,~\eqref{equation-ct-direct}
can be further simplified.
In particular,
for $N = 1024 = 32^2$,
we have
\begin{align}
\label{equation-ct-simplified}
\mathbf{X}
=
\operatorname{vec}
\left(
  \left[
  \mathbf{\Omega}_{32}
  \circ
  \left(
    \mathbf{F}_{32}
    \cdot
    (\operatorname{invvec}(\mathbf{x}))^\top
  \right)
  \right]
  \cdot
  \mathbf{F}_{32}^\top
\right)
.
\end{align}
The inner DFT call corresponds to row-wise transformation
of
$\operatorname{invvec}(\mathbf{x})$,
whereas the outer DFT performs
column-wise transformations on the resulting intermediate computation.
The formulation shown in~\eqref{equation-ct-simplified}
is the fundamental expression on
which the proposed approximations (eg. ADFTs) in this work
are based.

\section{Multiplierless 32-point ADFT}
\label{sec:32point}

In this section,
the adopted
multiplierless 32-point ADFT, first introduced in~\cite{DoraThesis, Madanayake2018},
is presented, and its complexity and error analysis are discussed . This is critical for understanding as the 1024-point  ADFT is realized using 32-point ADFT as the main building block.

\subsection{Matrix Representation}

The considered
32-point ADFT matrix denoted by $\hat{\mathbf{F}}_{32}$ can be
computed through a product of sparse matrices
whose real and imaginary parts of its coefficients
contains only $\pm 1$ entries.
Such simple arithmetic
leads to hardware designs
that can be realized with adders only.

To present the factorization of~$\hat{\mathbf{F}}_{32}$,
we need the auxiliary structures in eq. \eqref{Bt}- \eqref{eq:w7}, since the auxiliary factors are key for matrix factorization.
Let $\mathbf{B}_t$ be a~$t\times t$ real matrix given by
\begin{align}
\mathbf{B}_t &=
\begin{cases}
\left[
\begin{rsmallmatrix}
\mathbf{I}_{(t-1)/2} & & \bar{\mathbf{I}}_{(t-1)/2}\\
 & 1 & \\
\bar{\mathbf{I}}_{(t-1)/2} &  & -\mathbf{I}_{(t-1)/2}
\end{rsmallmatrix}
\right], & \text{if~$t$ is odd,}
\\
\\
\left[\begin{rsmallmatrix}
\mathbf{I}_{t/2} &  \bar{\mathbf{I}}_{t/2}\\
\bar{\mathbf{I}}_{t/2}  & -\mathbf{I}_{t/2}
\end{rsmallmatrix}
\right], & \text{if~$t$ is even,}
\end{cases} \label{Bt}
\end{align}
where~$\mathbf{I}_{k}$ and~$\bar{\mathbf{I}}_{k}$
being the identity and counter-identity matrix of order~${k}$, respectively.
Let also~$\mathbf{Z}_1$,~$\mathbf{Z}_2$, and~$\mathbf{Z}_3$ be the following matrices (for clarity, only the non-zero elements are shown):
\begin{align}
\mathbf{Z}_1=
\left[
\begin{smallmatrix}
1 & & & & & & & & & & & &1 & & & \\
 &1 & & & & & & & & & & & & & & \\
 & &1 & & & & & & & & & & & & & \\
 & & &1 & & & & & & & & & & & & \\
 & & & &1 & & & &1 & & & & & & & \\
 & & & & &1 & & & & & & & & & & \\
 & & & & & &1 & & & & & & & & & \\
 & & & & & & &1 & & & & & & & & \\
 & & & &1 & & & &-1 & & & & & & & \\
 & & & & & & & & &1 & & & & & & \\
 & & & & & & & & & &1 & & & & & \\
 & & & & & & & & & & &1 & & & & \\
1 & & & & & & & & & & & &-1 & & & \\
 & & & & & & & & & & & & &1 & & \\
 & & & & & & & & & & & & & &1 & \\
 & & & & & & & & & & & & & & &1 \\
\end{smallmatrix}
\right],
\end{align}
\begin{align}
\mathbf{Z}_2&=
\left[
\begin{smallmatrix}
1 & & & & & & & & & & & & & & & & \\
 &-1 & & & & & & & & & & & & & &1 & \\
 & &1 & & & & & & & & & & & & & & \\
 & & &1 & & & & & &1 & & & & & & & \\
 & & & &1 & &1 & &1 & & & & & & & & \\
 & & & & &1 & &1 & & & & & & & & & \\
 & & & &1 & &-1 & & & & & & & & & & \\
 & & & & &1 & &-1 & & & & & & & & & \\
 & & & &1 & & & &-1 & & & & & & & & \\
 & & &1 & & & & & &-1 & & & & & & & \\
 & & & & & & & & & &1 & & & & & & \\
 & & & & & & & & & & &1 & &1 & & & \\
 & & & & & & & & & & & &1 & &1 & &1 \\
 & & & & & & & & & & &1 & &-1 & & & \\
 & & & & & & & & & & & &1 & &-1 & & \\
 &1 & & & & & & & & & & & & & &1 & \\
 & & & & & & & & & & & &1 & & & &-1 \\
\end{smallmatrix}
\right],
\end{align}
and
\begin{align}
\mathbf{Z}_3&=
\left[
\begin{smallmatrix}
1 & & & &1 & &-1 & & & & & & & & \\
 &1 & & & & & & & & & & & & & \\
 & &1 &1 & & & & & & & & & & & \\
 & &1 &-1 & & & & & & & & & & & \\
1 & & & &-1 & & & & & & & & & & \\
 & & & & &1 & & & & & & & & & \\
1 & & & & & &1 & & & & & & & & \\
 & & & & & & &1 & & & & & & & \\
 & & & & & & & &1 & & & &1 & &-1 \\
 & & & & & & & & &1 & & & & & \\
 & & & & & & & & & &1 & & &1 & \\
 & & & & & & & & & & &1 & & & \\
 & & & & & & & &1 & & & &-1 & & \\
 & & & & & & & & & &1 & & &-1 & \\
 & & & & & & & &1 & & & & & &1 \\
\end{smallmatrix}
\right].
\end{align}

The 32-point ADFT
matrix
is
factorized into eight sparse matrices
$\mathbf{W}_k$,
for~$k = 0, 1, \ldots, 7$,
according to
\begin{align}
\label{eq:F32-fact}
\hat{\mathbf{F}}_{32} & =
\mathbf{W}_7
\cdot
\mathbf{W}_6
\cdot
\mathbf{W}_5
\cdot
\mathbf{W}_4
\cdot
\mathbf{W}_3
\cdot
\mathbf{W}_2
\cdot
\mathbf{W}_1
\cdot
\mathbf{W}_0
,
\end{align}
where
\begin{align}
	\mathbf{W}_0
	=
	\left[
	\begin{smallmatrix}
		\mathbf{B}_{17} & \\
		& \mathbf{B}_{15}
	\end{smallmatrix}
	\right]
	,
	\quad
	\mathbf{W}_1
	=
	\left[
	\begin{array}{c c}
		\mathbf{I}_{16} &
		\left[
		\begin{smallmatrix}
			0 & \\
			& \mathbf{I}_{15}
		\end{smallmatrix}
		\right]
		\\
		\left[
		\begin{smallmatrix}
			0 & \\
			& \mathbf{I}_{15}
		\end{smallmatrix}
		\right]
		& 	\left[	\begin{smallmatrix}
			1 & \\
			& -\mathbf{I}_{15}
		\end{smallmatrix}\right]
	\end{array}
	\right]
	,
\end{align}
\begin{align}
\mathbf{W}_2 &=
\left[
\begin{smallmatrix}
\mathbf{B}_{9} & & \\
& \mathbf{B}_{7} & \\
& & \mathbf{I}_{16}
\end{smallmatrix}
\right],
\quad
\mathbf{W}_3
=
\left[
\begin{smallmatrix}
\mathbf{B}_5 & &&&& &\\
 & 1&&&& &\\
& & \mathbf{B}_3&  &&& \\
& & & 1 & & &\\
& & & & \mathbf{B}_3 && \\
& & & & & \mathbf{B}_3 &\\
& & & & &  &
\mathbf{Z}_1
\end{smallmatrix}
\right],
\end{align}
\begin{align}
\mathbf{W}_4 &=
\left[
\begin{smallmatrix}
\mathbf{B}_3 & && &&& \\
& \mathbf{B}_2 && &&\\
&& \mathbf{B}_4 && &\\
&&& \mathbf{B}_4 && \\
&&&& \mathbf{B}_2 & \\
&&&&&
\mathbf{Z}_2
\end{smallmatrix}
\right]
,
\quad
\mathbf{W}_5 =
\left[
\begin{smallmatrix}
\mathbf{B}_2 & & \\
& \mathbf{I}_{15} &\\
&&
\mathbf{Z}_3
\end{smallmatrix}
\right],
\end{align}
\begin{align}
\mathbf{W}_6 =
\left[
\begin{smallmatrix}
\mathbf{I}_{16} & \\
&
\left[
\begin{smallmatrix}
1 & & & & & & & & & & & & &1 & & \\
 &1 & & & & & & &1 & & & & & & & \\
 & &-1 & & & & &1 & & & & & & & & \\
 & & &1 & & & & & & & & & & & & \\
 & & & &1 & & & & & & & & & & & \\
 & & & & &1 &1 & & & & & & & & & \\
 & & & & &1 &-1 & & & & & & & & & \\
 & &1 & & & & &1 & & & & & & & & \\
 &1 & & & & & & &-1 & & & & & & & \\
 & & & & & & & & &1 &1 & & & & & \\
 & & & & & & & & &1 &-1 & & & & & \\
 & & & & & & & & & & &1 & & & & \\
 & & & & & & & & & & & &1 & & &1 \\
1 & & & & & & & & & & & & &-1 & & \\
 & & & & & & & & & & & & & &1 & \\
 & & & & & & & & & & & &1 & & &-1 \\
\end{smallmatrix}
\right]
\end{smallmatrix}
\right],
\end{align}
and~$\mathbf{W}_7$ is given in~\eqref{eq:w7}.

\subsection{Arithmetic Complexity}
\label{sec:ADFT32-complexity}

In this section, we study the arithmetic complexity of the proposed ADFTs  by assuming execution is fully sequential. That is, we consider all algorithms execute on a sequential processor by utilizing a central processing unit (CPU) that furnishes arithmetic operations dictated by the particular algorithm. The execution time is proportional to the number of arithmetic operations, and in general, multiplication being more computationally intensive compared to addition, takes longer to execute. Therefore, the number of multiplications is the primary metric for quantification of arithmetic complexity.

The discussed 32-point ADFT has a null complexity of multiplications and no bit-shifting operations are required.
The only source of arithmetic complexity
is the number of additions in the factorization in~\eqref{eq:F32-fact}.
Considering complex inputs,
the matrices~$\mathbf{W}_0$,~$\mathbf{W}_1$, and~$\mathbf{W}_4$
require~60 real additions each, while the matrices~$\mathbf{W}_2$,~$\mathbf{W}_3$, and~$\mathbf{W}_5$
require~28 real additions each.
Similarly, the matrix~$\mathbf{W}_6$ requires 24
real
additions, while
the only complex matrix in the factorization,
$\mathbf{W}_7$,
requires~60 real additions.
In total, the transform~$\hat{\mathbf{F}}_{32}$ thus requires~348 real additions and no bit-shifting.
By comparison,
the Cooley-Tukey radix-2 algorithm
requires~88 real multiplications
and~408 real additions~\cite{blahut2010fast,Duhamel1984}. In contrast, the approach to represent a 32-point DFT using~\eqref{equation-dft-def-n-squared} and~\eqref{equation-ct-simplified} offers $1/8$ the number of additions with no multiplications as compared to the 88 multiplications needed by Cooley-Tukey.

\subsection{Error Analysis}

\begin{figure}
\centering
\psfrag{-pi}[][][0.7]{$-\pi$}
\psfrag{-pi/2}[][][0.7]{$-\pi/2$}
\psfrag{pi/2}[][][0.7]{$\pi/2$}
\psfrag{pi}[][][0.7]{$\pi$}
\subfigure[32-point DFT]{\label{fig:filter_32DFT}\epsfig{file=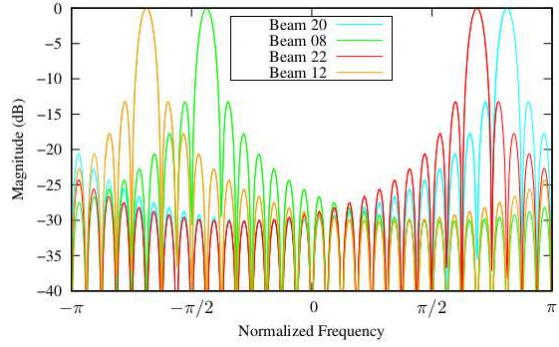, scale=0.5}}\\
\subfigure[32-point ADFT]{\label{fig:filter_32ADFT}\epsfig{file=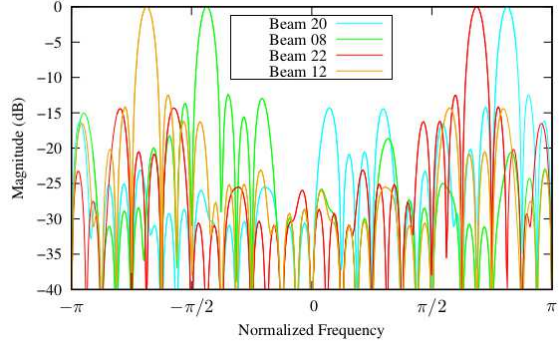, scale=0.5}}\\
\subfigure[Error]{\label{fig:filter_32ERR}\epsfig{file=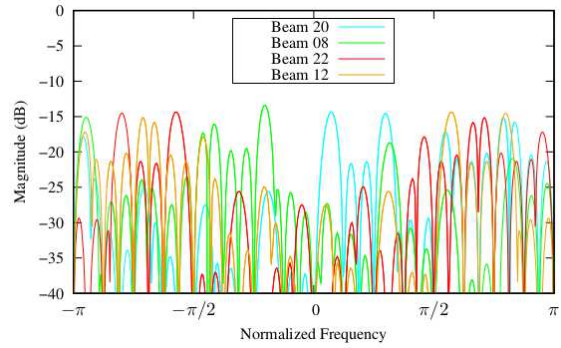, scale=0.5}}
\caption{Magnitude of the filter-bank responses for
(a)~the exact 32-point DFT,
(b)~the 32-point ADFT
and
(c)~error of the ADFT response
for the least performing rows.
\label{32Plots}}
\end{figure}

The rows of a linear transform matrix
can be understood as a finite impulse response (FIR)
filter bank~\cite{Oppenheim2009}. Savings of computation and exploitation of sparsity gives rise to slightly inaccurate representations of the frequency response of the filter bank.
Thus
we can assess
how close
the filter bank implied by the proposed (in eq. \eqref{eq:F32-fact}) ADFT approximations
are to the rows of exact DFT matrix.
The filter bank frequency responses
for four of the bins of the 32-point DFT, 32-point ADFT,
and the corresponding error plots
are shown in Fig.~\ref{32Plots}.
The four bins shown are the ones corresponding the
the rows of the 32-point ADFT
that performs the worst in terms of
frequency response;
thus they can be understood as worst-case scenarios.
The figure shows that the 32-point ADFT is ``close enough'' to the exact DFT to be useful in many practical applications,
especially for wireless
communications and software-defined radio (SDR) where antenna pattern, noise and semiconductor process variations induce some errors themselves: its error level of about $-10$~dB compared to the main lobe of exact DFT's filterbank response is
within the margin of error of such systems (which include both electronics and electromagnetics).

\begin{figure*}[!t]
\normalsize
\begin{align}
\label{eq:w7}
\mathbf{W}_7 &=
\left[
\begin{smallmatrix}
1 & &  & &  & & & &  & &  &  &  & & & &  &  &  & &  & & &  & &  &  &  &  &  & &  \\
 &  &  & &  &  & & &  & & & &  & &  &  & & &  &-j &  &  &  &  & & & &1 & &  &  & \\
  & &  &  & & &1 & & & & -j & & &  & & &  &  &  &  &  &  & &  &  &  &  & &  & &  &  \\
 &  & &  & &  & & & & &  & & & & &  &  & &  &  &  &  & & -j &  &  &  &  &-1 & &  & \\
  & &  &1 & & &  &  &  &  &  & & &j & &  &  & &  & & &  & &  & & &  &  & & &  &  \\
  &  &  & &  & &  & & & & & &  &  & & &  & -j &  &  &  & & & &  &1 &  &  & &  & &  \\
  & & & &  &-1 & &  & &-j &  & & & & & & &  &  & & &  &  & & &  &  &  & &  &  & \\
  & & &  &  &  &  &  & & &  & & & &  & &-1 &  &  & &  &  & -j & & &  &  &  &  &  &  & \\
  & &1 & & & &  &  & & &  & & & &  &-j & &  &  & &  &  & & &  &  &  &  &  & &  &  \\
  &  &  & &  &  & & &  & & &  &  & &  &  & &  & & &  & -j & & & & &  & & &-1 &  & \\
  & & & &  &  & & &1 &  &  &  &-j & &  & & &  &  & & &  & & & &  & &  &  &  &  &  \\
  &  &  & &  &  & &  &  & &  & & & &  &  & & &  & &  &  & & &-j &  &-1 & &  &  &  & \\
  & &  & &-1 & & &  &  &  &  & & &  & j & & &  & & &  &  & & & &  &  & &  & &  & \\
  &  & & &  &  &  &  & & &  & & &  & &  & & & -j & & & & & & &  &  &  & & &  &-1 \\
  &  & & &  & & &1 & &  &  &j & & &  & &  &  & & & & & & & &  &  & &  & &  & \\
  & &  & &  &  & &  & & &  & & &  &  &  & & &  &  &-j & & &  & &  &  & & &  &-1 & \\
  &1 &  &  &  & & &  &  &  &  & & & &  & & &  &  &  & & & & & &  &  &  & &  &  & \\
  & &  & &  &  & &  & & &  & & &  &  &  & & &  &  &j & & &  & &  &  & & &  &-1 & \\
  &  & & &  & & &1 & &  &  &-j & & &  & &  &  & & & & & & & &  &  & &  & &  & \\
  &  & & &  &  &  &  & & &  & & &  & &  & & & j & & & & & & &  &  &  & & &  &-1 \\
  & &  & &-1 & & &  &  &  &  & & &  & -j & & &  & & &  &  & & & &  &  & &  & &  & \\
  &  &  & &  &  & &  &  & &  & & & &  &  & & &  & &  &  & & &j &  &-1 & &  &  &  & \\
  & & & &  &  & & &1 &  &  &  &j & &  & & &  &  & & &  & & & &  & &  &  &  &  &  \\
  &  &  & &  &  & & &  & & &  &  & &  &  & &  & & &  & j & & & & &  & & &-1 &  & \\
  & &1 & & & &  &  & & &  & & & &  &j & &  &  & &  &  & & &  &  &  &  &  & &  &  \\
  & & &  &  &  &  &  & & &  & & & &  & &-1 &  &  & &  &  & j & & &  &  &  &  &  &  & \\
  & & & &  &-1 & &  & &j &  & & & & & & &  &  & & &  &  & & &  &  &  & &  &  & \\
  &  &  & &  & &  & & & & & &  &  & & &  & j &  &  &  & & & &  &1 &  &  & &  & &  \\
  & &  &1 & & &  &  &  &  &  & & &-j & &  &  & &  & & &  & &  & & &  &  & & &  &  \\
 &  & &  & &  & & & & &  & & & & &  &  & &  &  &  &  & & j &  &  &  &  &-1 & &  & \\
  & &  &  & & &1 & & & & j & & &  & & &  &  &  &  &  &  & &  &  &  &  & &  & &  &  \\
 &  &  & &  &  & & &  & & & &  & &  &  & & &  &j &  &  &  &  & & & &1 & &  &  & \\
\end{smallmatrix}
\right].
\end{align}
\hrulefill
\end{figure*}%

\section{Approximations
for the 1024-point DFT}\label{s:1024}

\subsection{Approximation Methodology}

As section \ref{sec:ADFT32-complexity} illustrated the ADFT for $N = 32$, here we exploit the square of $N = 32$ to create a family of ADFTs for $N=1024$.
Motivated by the promising results achieved for 32-point ADFT, we will extend the approximation to 1024-point case using the mathematics described in section \ref{sec:DFT-N-squared}. Here, we propose three ADFT algorithms which have small
deviations of their filter bank responses when
compared to the DFT.
We assume that the applications at hand will be tolerant
of the given deviations of frequency response,
and that such deviations will be a small price to pay in
exchange for the significantly smaller circuit realizations
and power consumption over traditional fixed-point FFTs.
It should be noted that
the
implementation of such approximate methods
is not constrained by
the minimum theoretical bounds of multiplicative complexity~\cite{heideman},
that apply to the exact DFT.
Indeed
the proposed algorithms
are not in fact calculating the DFT,
but furnishing
approximations that are deemed reasonable for most high-speed digital-RF
applications.

Based on~\eqref{equation-ct-simplified},
we propose the replacement
of
the exact 32-point DFT~$\mathbf{F}_{32}$
by
the 32-point ADFT proposed in~\cite{DoraThesis}.
Therefore,
a suite of approximations
for the DFT computation emerges.
We propose three different algorithms based on the position of ADFT matrix in the derivation:
\begin{itemize}
\item
\textbf{Algorithm 1: ADFT-ADFT.}
Substitute both row- and column-wise 32-point DFT
$\mathbf{F}_{32}$
with
the multiplierless 32-point ADFT
$\hat{\mathbf{F}}_{32}$;

\item
\textbf{Algorithm 2: Hybrid ADFT-DFT.}
Replace only the row-wise 32-point FFTs with
the multiplierless 32-point ADFT in Section~\ref{sec:32point}
leaving column-wise DFTs exact, and;

\item
\textbf{Algorithm 3: Hybrid DFT-ADFT.}
Replace only the column-wise 32-point FFTs with
the multiplierless 32-point ADFT in Section~\ref{sec:32point}
leaving row-wise DFTs exact.
\end{itemize}
Let~$\hat{\mathbf{X}}_i$ for~$i = 1, 2, 3$ denote
approximations for~$\mathbf{X}$ given by
Algorithm~1, Algorithm~2, and Algorithm~3, respectively.
Thus
we have mathematically:
\begin{align}
\hat{\mathbf{X}}_1
&
=
\operatorname{vec}
\left(
  \left[
  \mathbf{\Omega}_{32}
  \circ
  \left(
    \hat{\mathbf{F}}_{32}
    \cdot
    (\operatorname{invvec}(\mathbf{x}))^\top
  \right)
  \right]
  \cdot
  \hat{\mathbf{F}}_{32}^\top
\right)
,
\label{eq:alg1}
\end{align}

\begin{align}
\hat{\mathbf{X}}_2
&
=
\operatorname{vec}
\left(
  \left[
  \mathbf{\Omega}_{32}
  \circ
  \left(
    \hat{\mathbf{F}}_{32}
    \cdot
    (\operatorname{invvec}(\mathbf{x}))^\top
  \right)
  \right]
  \cdot
  \mathbf{F}_{32}^\top
\right)
,
\label{eq:alg2}
\end{align}
and
\begin{align}
\hat{\mathbf{X}}_3
&
=
\operatorname{vec}
\left(
  \left[
  \mathbf{\Omega}_{32}
  \circ
  \left(
    \mathbf{F}_{32}
    \cdot
    (\operatorname{invvec}(\mathbf{x}))^\top
  \right)
  \right]
  \cdot
  \hat{\mathbf{F}}_{32}^\top
\right)
.
\label{eq:alg3}
\end{align}
The above
combinations of ADFT and DFT
yield low-complexity
approximations
for the 1024-point DFT,
which---due to its relatively large block length---is
a computationally intractable task
via usual direct numerical search methods.
Algorithms 1,~2, and~3
have considerably different computational complexities
and performance trade-offs,
as discussed in subsection \ref{arith-complex}.

\subsection{Arithmetic Complexity}\label{arith-complex}

\subsubsection{Twiddle-factor Matrix}

In the three proposed algorithms,
only the DFT computation
$\mathbf{F}_{32}$
is subject to an approximation;
the twiddle-factor matrix~$\mathbf{\Omega}_{32}$
is left unaltered in its exact form (cf.~\eqref{equation-ct-simplified}).
Therefore,
a minimum number of multiplications
remains due to~$\mathbf{\Omega}_{32}$.
Considering only the nontrivial
multiplications,
the twiddle-factor matrix
requires
961 complex multiplications,
which
translate
into
2883 real multiplications
and
2883 real additions.
The arithmetic complexity assumes sequential operation in a CPU. This parameter will be used in the arithmetic complexity calculations for each of the three algorithms.

\subsubsection{Algorithm 1}

Here the only source of multiplicative complexity
are the twiddle factors
in between the row- and column-wise 32-point ADFT blocks.
Since the 32-point ADFT
requires 348~additions
and
it is called 64~times,
it contributes~$64 \times 348 = 22272$ real additions
to the overall arithmetic complexity of Algorithm~1.
The resulting arithmetic costs are:
2883~real multiplications
and
$2883 + 22272 = 25155$~additions.

\subsubsection{Algorithm 2}

Here
multiplicative costs
stem
from
the twiddle factors
and
the column-wise 32-point exact DFT.
The column-wise exact DFT is computed using
the Cooley-Tukey radix-2 FFT~\cite{blahut2010fast,Duhamel1984}
(see~Section~\ref{sec:ADFT32-complexity}).
Since this algorithm requires~$32$ calls
to the exact 32-point DFT
and~32
calls to the 32-point ADFT,
we have a total of
$(32\times 88) + 2883 = 5699$ real multiplications
and
$(32\times 408) + (32 \times 348) + 2883 = 27075$
real additions.

\subsubsection{Algorithm 3}

Here the operation count
follows the same rationale as for Algorithm~2,
with the difference
that the roles of the row and column-wise
transforms are swapped.
Therefore,
Algorithms~2 and 3 have the same arithmetic costs.
The arithmetic complexity
of the proposed methods
is summarized in Table~\ref{tab:arithmetic-complexity}.

\begin{table}
\centering
\caption{Real arithmetic complexity
for the exact 1024-point DFT
and
for the proposed approximations}
\centering
\begin{tabular}{l@{\quad}c@{\quad}c@{\quad}}
\toprule
Algorithm & Real Mult. & Real Additions  \\\midrule
Cooley-Tukey Radix-2 FFT~\cite[p.~76]{blahut2010fast} & 10,248 & 30,728  \\\midrule
Split-Radix FFT~\cite{Duhamel1984}& 7,172 & 27,652  \\\midrule
Winograd FFT~\cite{Winograd1980} & 10,248 & 30,728  \\\midrule
Proposed Algorithm~1 & 2,883 & 25,155\\\midrule
Proposed Algorithm~2 & 5,699 & 27,075\\\midrule
Proposed Algorithm~3 & 5,699 & 27,075\\
\bottomrule
\end{tabular}
\label{tab:arithmetic-complexity}
\end{table}

\subsection{Performance of the Proposed Approximations}
\label{sec:comp-accuracy}
\begin{figure}
\centering
\psfrag{XXXX}[][][0.4]{ZZZ}
\psfrag{Quant1XXXX}[][][0.4]{First Quartile}
\psfrag{Quant2XXXX}[][][0.4]{Second Quartile}
\psfrag{Quant3XXXX}[][][0.4]{Third Quartile}
\psfrag{Upper Bound}[][][0.4]{Upper Bound}
\psfrag{Lower Bound}[][][0.4]{Lower Bound}
\psfrag{-pi/4}[][][0.7]{$-\pi/4$}
\psfrag{-pi/8}[][][0.7]{$-\pi/8$}
\psfrag{pi/8}[][][0.7]{$\pi/8$}
\psfrag{pi/4}[][][0.7]{$\pi/4$}

\subfigure[Algorithm~1]{\label{fig:FIR_error_ADFT1}\epsfig{file=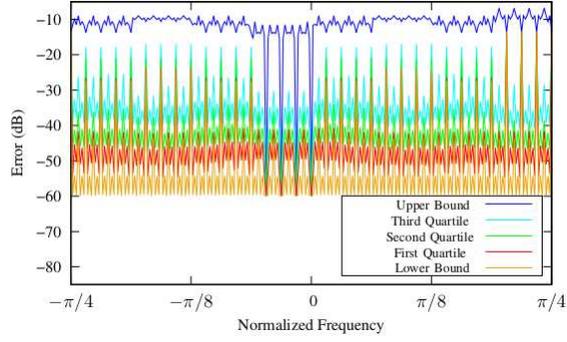, scale=0.5}}\\
\subfigure[Algorithm~2]{\label{fig:FIR_error_ADFT2}\epsfig{file=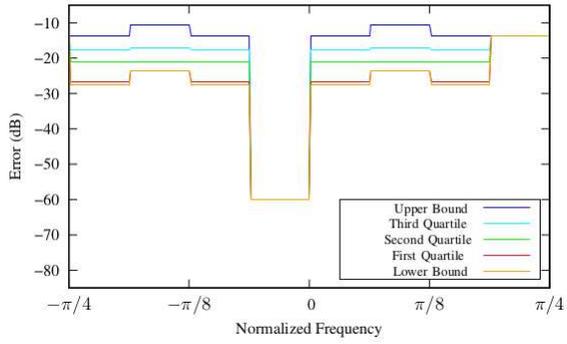, scale=0.5}}\\
\subfigure[Algorithm~3]{\label{fig:FIR_error_ADFT3}\epsfig{file=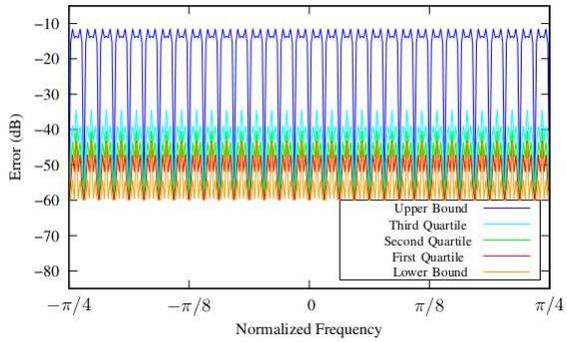, scale=0.5}}

\caption{
Log-magnitude error of the frequency response of the rows of the proposed approximations.
The errors are bounded to $-60$~dB for displaying purposes.}

\label{fig:FIR_error}
\end{figure}

Considering
the frequency response error
expressed in log-magnitude units, Fig.~\ref{fig:FIR_error}
shows
(i)~the upper and lower envelopes
and
(ii)~the first, second, and third quartiles
of the error
resulting from
the proposed approximate filter banks~\cite{Suarez2014, Kulasekera2015}.
For ease of visual inspection,
we show only the normalized frequencies on the interval~$[-\pi/4, \pi/4]$.
The error of the frequency response for the remaining parts of the interval~$[-\pi, \pi]$
are just a repetition of the plots in Fig.~\ref{fig:FIR_error}.

Note that the
three approximations
resulting from Algorithm~1, Algorithm~2, and Algorithm~3
have distinct frequency responses.
Fig.~\ref{fig:FIR_error} indicates that
the Algorithm~1 is the one presenting
the largest
deviation for the main lobe from the exact DFT.
This is expected given
that the transform resulting from
Algorithm~1 is obtained through the substitution of
both the row- and column-wise DFT block
by the discussed approximate 32-point DFT.
This qualitative analysis is confirmed once we calculate
the errors in the frequency responses of the rows
of the three proposed approximations.
Table~\ref{tab:error_metrics}
displays
the minimum (nonzero), mean, and maximum
for the squared magnitude of these
errors.
Notice in Fig. \ref{fig:FIR_error} that the transform resulting from Algorithm~1
has the highest deviations from
the expected frequency response for its rows with range of~$5$~dB compared to the filter bank response of the exact DFT matrix.
In Table~\ref{tab:error_metrics},
we also show the worst-case side lobe in dB for each of the
transforms.
All  transforms considered here
possess
a low worst-case side lobe on the order of~$-12$~dB.

Noise rejection of the proposed ADFTs can be evaluated by means of its SNR improvement per frequency bin.
The noise present from the antenna array can be modeled  as additive white Gaussian noise (AWGN) with zero mean and variance~$\sigma^2$. The AWGN present in each frequency bin is $\sigma^2/N.$
For narrowband (monochromatic) plane wave received by the array, the input signals to both the DFT and the three ADFT algorithms
follows~$\exp(j2\pi n k / N)$ for~$n,k = 0, 1, \ldots, N-1$, where~$k$ represents the DFT/ADFT bin number (corresponding to specific spatial frequencies related to the direction of propagation of each wave) and~$n$ is the antenna number in the ULA.
The monochromatic signal having frequency $\exp(j2\pi k/N)$ for bin $k$ has its SNR improved
by~$10 \operatorname{log}_{10} (N)$, which is $30.1~$dB for the 1024 point DFT.
This is the best case SNR improvement per bin for the DFTs.
The adoption of various ADFTs in place of the DFT causes a loss of SNR performance observed as a hit in the SNR per bin. Let the reduction in SNR for bin $k$ be denoted
$\Delta \gamma_{x}$ where $x\in \{1,2,3 \}$ are for the three proposed approximation algorithms.

The worst-case SNR degradation for the ADFTs obtained through simulations with~$10^5$ replicates for finding the ensemble average for each bin of the ADFTs are shown in Table~\ref{tab:error_metrics}.
The SNR degradation shows that Algorithm~1 has the largest worst-case degradation of SNR compared to the DFT ($\Delta \gamma_1< 0.9~$dB).
There is no significant difference between Algorithm~2 and Algorithm~3 in terms of SNR degradation and has worst-case ($\Delta \gamma_{2,3}<0.4~$dB). The reduction in SNR of the three ADFTs compared to SNR of DFTs can be compensated by adding  $\le 0.9~$ dB of additional transmit power and antenna gain at the transmit side.
\begin{figure*}
\centering
\psfrag{0}[][][0.7]{$0$}
\psfrag{256}[][][0.7]{$256$}
\psfrag{512}[][][0.7]{$512$}
\psfrag{768}[][][0.7]{$768$}
\psfrag{1023}[][][0.7]{$1023$}
\psfrag{Beam}[][][0.7]{$k$th beam}
\subfigure[DFT]{\label{fig:DFT_SNR}\epsfig{file=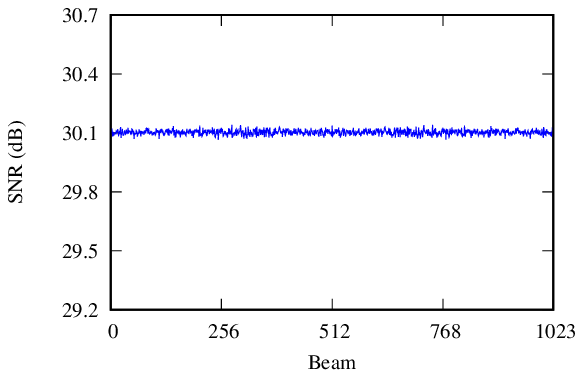}}
\subfigure[Algorithm~1]{\label{fig:ADFT1_SNR}\epsfig{file=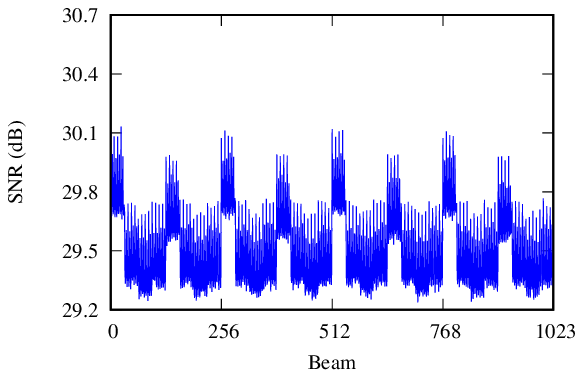}}
\subfigure[Algorithm~2]{\label{fig:ADFT2_SNR}\epsfig{file=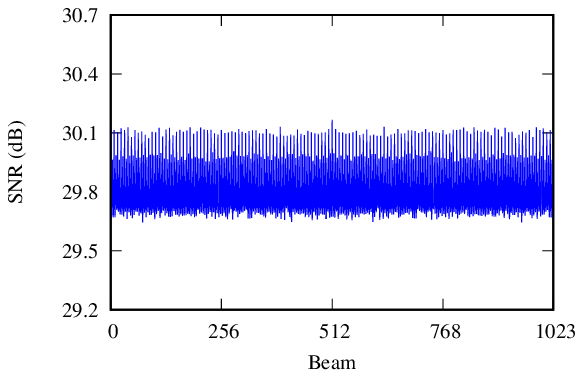}}
\subfigure[Algorithm~3]{\label{fig:ADFT3_SNR}\epsfig{file=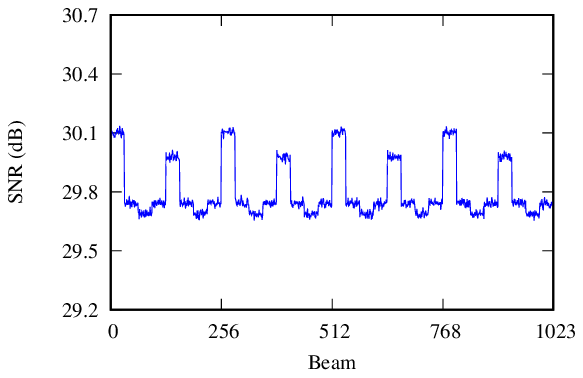}}
\caption{
The SNR for DFT and proposed Algorithm~1, Algorithm~2, and Algorithm~3.
The SNR for the proposed algorithms is no lower than~$29.2$ dB, against~$30.1$ dB for the DFT.}
\label{fig:SNR_simulation}
\end{figure*}
Fig.~\ref{fig:SNR_simulation}
shows the SNR plot for each of the beams of the DFT
and the three proposed approximations.
Notice that no approximation has an SNR lower than~$29.2$~dB in any of the bins, demonstrating that the SNR degradation is $\le 0.9$~dB compared to the DFT where the SNR improves by~$30.1$ dB for every bin.

\begin{table*}
\centering
\caption{Performance statistics of the proposed approximations: frequency response magnitude, worst-case side lobe level and SNR degradation}
\label{tab:error_metrics}
\begin{tabular}{c@{\quad}c@{\quad}c@{\quad}c@{\quad}c@{\quad}c@{\quad}}
\toprule
\multirow{2}{*}{Transform} & \multicolumn{3}{c}{Error Magnitude of Rows} & Worst Side & SNR \\\cmidrule{2-4}
 & {Min (dB)} & {Mean (dB)}& {Max (dB)} & Lobe (dB ) & degradation (dB) \\\midrule
Algorithm~1 & $-10.7$ & $-5.5$& $-4.4$ & $-12.8$ & $-0.6$\\\midrule
Algorithm~2 & $-10.7$ & $-9.9$ & $-9.0$ & $-12.8$& $-0.3$ \\\midrule
Algorithm~3 & $-10.7$ & $-9.9$ & $-9.0$ & $-12.9$& $-0.3$\\\bottomrule
\end{tabular}
\end{table*}

\section{Digital VLSI Realization}\label{s:vlsi}

\begin{figure}
\centering
\includegraphics[scale=0.75]{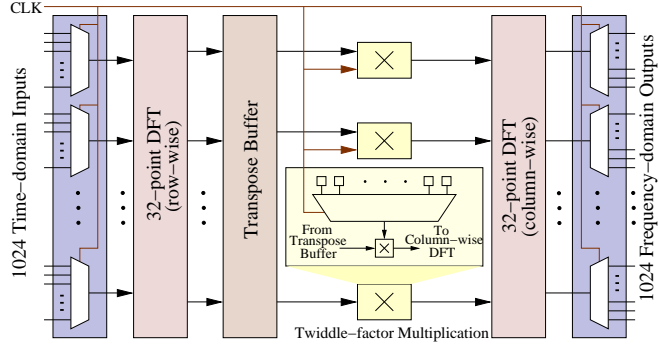}
\caption{Signal flow graph showing the VLSI architecture to be modified
for the proposed architecture based on the selected approximation.
\emph{Algorithm 1:} Replacement of both 32-point DFTs with 32-point ADFT blocks.
\emph{Algorithm 2:} Replacement of only row-wise 32-point DFT with 32-point ADFT blocks leaving column-wise DFT exact.
\emph{Algorithm 3:} Replacement of column-wise 32-point FFT with 32-point ADFT blocks leaving row-wise DFT exact.
\label{algorithm_block}}

\end{figure}

Next, we explore digital VLSI realizations
of the three ADFT approaches outlined in~\eqref{eq:alg1}, \eqref{eq:alg2} and~\eqref{eq:alg2}
using a time-multiplexed approach. Traditionally, arithmetic complexity amounts of counts of
both multiplication operations and addition operations. However, for semi-parallelized hardware implementations on
VLSI platforms, the existence of parallel sub-systems offers a trade-off between circuit complexity and algorithm execution speed
as described by Amdahl's Law \cite{Amdahl1967}. The proposed algorithms are based on radix-32 SFGs, which imply the sequential nature is limited to 1024-point algorithm completion every 32 clock cycles. The radix-32 SFG allows re-use of ADFT and DFT cores, and twiddle-factor cores, using time-multiplexing up to 32 levels. The use of time-multiplexed operations leads to the generalization of the multiplier structures that do not distinguish trivial multiplications by $0,1,-1.$ Therefore, the number of multiplications for the twiddle factor block is 1024 complex multiplications  (compare to 961 complex multiplications for a sequential algorithm that can ignore trivial multiplications). However, radix-32 approach allows time-multiplexing of 32 parallel complex multipliers for achieving the twiddle factor matrix, leading to circuit complexity of 96 real multipliers, and 160 real adders/subtractors, in the twiddle factor block.
To distinguish the mathematical operations from its
physical realization,
hereafter
we refer to the circuit implementation of the
selected 32-point DFT and ADFT, respectively,
as
\texttt{DFT32} and \texttt{ADFT32} cores.
Also, the
digital VLSI hardware for the 1024-point exact DFT and each
of the 1024-point approximations resulting from
Algorithm~1, Algorithm~2, and Algorithm~3
are referred to as
the~\texttt{DFT1024},
\texttt{ADFT1024\_1},
\texttt{ADFT1024\_2},
and~\texttt{ADFT1024\_3} cores,
respectively.
Fig.~\ref{algorithm_block}
shows the overall
architecture
of the
\texttt{DFT1024}
with
the
\texttt{DFT32} cores.
We focus on the design of the
\texttt{ADFT1024\_1} core.
Because
this design can be easily extended to the other
cores,
the description of the
\texttt{ADFT1024\_2} (Algorithm~2)
and
\texttt{ADFT1024\_3} (Algorithm~3)
cores
is omitted for brevity.

The
core~\texttt{ADFT1024\_1}
is a radix-32 unit and therefore processes
an input signal block
of 1024~time-domain samples in 32~clock cycles.
Each
signal
block
consists of 32~rows of adjacent time-domain samples in 32~columns.
The first~\texttt{ADFT32}
block
sequentially
computes the
32-point ADFT of
each row,
which are given by:
$x[k], x[32 + k], x[2\times 32 + k], \ldots, x[31\times 32 + k]$,
for
$k = 0, 1, \ldots, 31$.
Sampled values in the intermediate frequency (IF) domain are passed
to the transpose buffer,
which realizes the matrix transposition operation
in digital~\mbox{VLSI} hardware,
while operating in-step with the system clock.
One complete matrix transpose operation is
achieved every 32~clock cycles.
The transpose buffer feeds the second
time-multiplexed~\texttt{ADFT32} after suitable
twiddle factors have been applied, which in turn,
furnishes the desired 1024-point ADFT values.
In order to minimize the chances of overflow,
the second
time-multiplexed~\texttt{ADFT32} block in Fig.~\ref{algorithm_block}
uses a larger wordlength by one bit than the first
time-multiplexed~\texttt{ADFT32} block.
Use of a larger word length accommodates for the arithmetic operations that are
carried on the first
time-multiplexed~\texttt{ADFT32}
and the twiddle factors.

\subsection{Transpose Buffer and Twiddle Factors}

\begin{figure}
\centering
\includegraphics[scale=0.8]{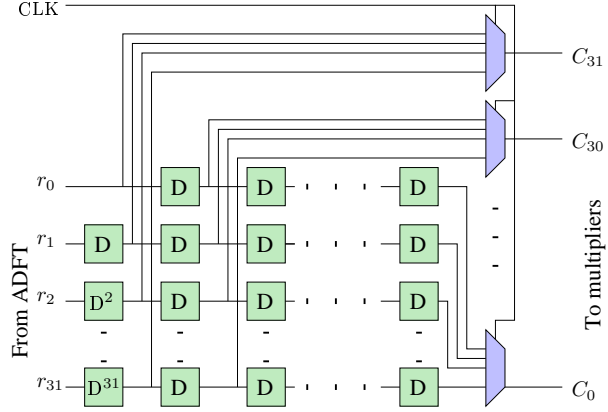}
\caption{Schematic diagram of the transpose buffer.}
\label{trans_buf}
\end{figure}

The transpose buffer shown in Fig.~\ref{trans_buf}
consists of a mesh of 1024 delays and 32 parallel multiplexers,
each of them possessing 32 inputs.
The transpose buffer block generates the transpose of the first set of frequency bins.
The transposition allows the column-wise DFT computation required in eq. \eqref{eq:alg1}, \eqref{eq:alg2} and \eqref{eq:alg3}.

Twiddle-factor multiplication count consists of
961 non-trivial complex multiplications spread over 32 clock cycles. However,
these are implemented using 32 parallel complex multipliers, which each consume 3 real multipliers and 5 adders (Gauss Algorithm
for complex multiplication).
The twiddle factor block therefore
furnishes the only multiplications
present
in \texttt{ADFT1024\_1}, which results from Algorithm~1 in a radix-32 hardware realization.
Each of the column bins (after the transpose buffer) undergoes
a multiplication by
$\omega_{1024}^{m\cdot n}$,
where~$ 0 \leq m \leq 31$ and~$0 \leq n \leq 31$.
Therefore,
the precision of the twiddle-factor multipliers
plays
a critical role in the final~area~$A$,
area-time~$AT$, and
area-time-squared~($AT^2$)~metrics.
In this paper, we have set the twiddle-factor precision
level to be equal to the system word size of the inputs to
the~\texttt{ADFT1024\_1} core, which is a design parameter and the choice of lower precision
levels in the twiddle factors would result
in improvements in the~VLSI metrics
for all three proposed algorithms.
In a sense,
hardware designed with such conservative parameters
can be thought of as worst-case benchmark,
with more coarsely quantified twiddle
factors leading to even better improvements
in area, area-time, and area-time-squared metrics.

\subsection{Circuit Complexity}

All circuits operate for 32 clock cycles to produce one 1024-point transform.
Complex multiplication is realized using 3 real multiplier circuits and 5 real adder circuits following the Gauss multiplication algorithm. The twiddle-factor matrix based on Gauss multiplication~$\mathbf{\Omega}_{32}$
in~\eqref{eq:alg1}
therefore requires 96~real multiplier circuits and 160 adders circuits. This block is common to all four 1024-point algorithms.

\subsubsection{\texttt{ADFT1024\_1} core}

Each \texttt{ADFT32} requires
348~adders/subtractors and no multipliers.

As shown in Fig.~\ref{algorithm_block},
the proposed radix-32 time-multiplexed architecture for Algorithm~1 uses
two~\texttt{ADFT32} cores.
Thus,~\texttt{ADFT1024\_1} has an overall circuit complexity of
$348*2+160=856$~adders/subtractor circuits
and
$96$~real multiplier circuits.

\subsubsection{\texttt{ADFT1024\_2} core}
In~\texttt{ADFT1024\_2}, the  row-wise DFT block is substituted by
the  ADFT32 block.
The~\texttt{DFT32} requires a total
$78$~real multiplier circuits and $398$ adder circuits.
Because the~\texttt{ADFT32} requires~$348$ adder circuits but no multipliers,
we have an overall circuit complexity of
$398+160+348= 906$~adder circuits
and
$96+78=174$~multipliers~\texttt{ADFT1024\_2}.

\subsubsection{\texttt{ADFT1024\_3} core}

The circuit complexity  for the
Algorithm~3 is the same as for \texttt{ADFT1024\_2}.
The only change is in the placement of the elements in the
architectural level.

The 1024-point DFT (denoted DFT1024) obtained by using two DFT32 cores for row- and column-wise FFTs would require
$78*2+96=252$ real multiplier circuits and $398*2+160=796+160=956$ adder circuits. This is our reference radix-32 FFT circuit for baselining the circuit complexities of the proposed ADFT1024 algorithms.

The circuit complexities for
the proposed designs as well as~\texttt{DFT1024}
are presented in Table~\ref{tab:circuit_complexity}.

\begin{table}
\centering
\caption{Circuit complexity for the proposed architectures and the 1024-point DFT}
\label{tab:circuit_complexity}
\begin{tabular}{c@{\quad}c@{\quad}c@{\quad}}
\toprule
Design & Multiplier Circuits & Adder Circuits\\\midrule
$DFT1024$ & $252$ & $959$\\\midrule
$ADFT1024\_1$ & $96$ & $856$\\\midrule
$ADFT1024\_2$ & $174$ & $906$\\\midrule
$ADFT1024\_3$ & $174$ & $906$\\\bottomrule
\end{tabular}
\end{table}

\subsection{ASIC Synthesis and Place-Route Results: 45nm CMOS}
\label{s:asic}

The proposed architectures were implemented on MATLAB
Simulink using Xilinx libraries and then mapped to 45-nm complementary
metal-oxide semiconductor \mbox{(CMOS)} technology cells (synthesis only).
Each of the designs consists of three main hardware
components---first 32-point transform block,
transpose buffer with twiddle-factor multiplication block,
and
second 32-point transform block.
The complexity of each 32-point transform block
core depends on its
corresponding input word length.
Key quantitative
measurements of performance
for each 32-point transform block core and transpose
buffer with twiddle-factor multiplications are
shown in Table~\ref{comp_table}.
In Table~\ref{algo_table},
we list the hardware
implementation metrics for
\texttt{ADFT1024\_1},~\texttt{ADFT1024\_2}, and~\texttt{ADFT1024\_3}.
Metrics for the \texttt{DFT1024} core were included as reference values.

\begin{table*}[]
\caption{Key quantitative measurements of performance in digital 45~nm CMOS VLSI for each DFT core and transpose buffer with twiddle factor multiplications\label{comp_table}}
\centering
\begin{tabular}{l|c|c|l|c|c|c|l}
\hline
\multicolumn{1}{c|}{\multirow{2}{*}{\begin{tabular}[c]{@{}c@{}}Performance \\ Metric\end{tabular}}} & \multicolumn{3}{c|}{Row-wise transform} & {\multirow{2}{*}{\begin{tabular}[c]{@{}c@{}}Transpose buffer\\ with multipliers\end{tabular}}} & \multicolumn{3}{c}{Column-wise transform} \\ \cline{2-4}  \cline{6-8}
\multicolumn{1}{c|}{} & \texttt{DFT32} & \texttt{ADFT32} & \multicolumn{1}{c|}{Change} &  & \texttt{DFT32} & \texttt{ADFT32} & \multicolumn{1}{c}{Change} \\ \hline
Area, $A$ (mm$^2)$ & 4.4 & 1.0 & 76.9 $\%\downarrow$ & 6.9 & 6.8 & 1.5 & 78.4 $\%\downarrow$ \\ \hline
\begin{tabular}[c]{@{}l@{}}Critical Path Delay, $T$ \\ (ns)\end{tabular} & 1.9 & 1.8 & 3.1 $\%\downarrow$ & 1.8 & 1.9 & 1.8 & 3.1 $\%\downarrow$ \\ \hline
\begin{tabular}[c]{@{}l@{}}Frequency, $F_\text{max}$\\ (GHz)\end{tabular} & 0.5 & 0.6 & 3.2 $\%\uparrow$ & 0.6 & 0.5 & 0.6 & 3.2 $\%\uparrow$ \\ \hline
\begin{tabular}[c]{@{}l@{}}$AT$\\ (mm$^2$ns)\end{tabular} & 8.3 & 1.9 & 77.7 $\%\downarrow$ & 12.6 & 12.9 & 2.7 & 79.1 $\%\downarrow$ \\ \hline
\begin{tabular}[c]{@{}l@{}}$AT^2$\\ (mm$^2$ns$^2$)\end{tabular} & 15.7 & 3.4 & 78.4 $\%\downarrow$ & 22.9 & 24.2 & 4.9 & 79.8 $\%\downarrow$ \\ \hline
\begin{tabular}[c]{@{}l@{}}Dynamic Power, $D_p$\\ (mW/GHz)\end{tabular} & 10.1 & 2.0 & 80.1 $\%\downarrow$ & 10.4 & 24.3 & 2.8 & 88.5 $\%\downarrow$ \\ \hline
\end{tabular}
\end{table*}

\begin{table*}[]
\caption{Key quantitative measurements of performance in digital 45~nm CMOS VLSI  for each algorithm\label{algo_table} }
\centering
\begin{tabular}{l|c|c|c|c|c|c|c}
\hline
\multicolumn{1}{c|}{\multirow{2}{*}{\begin{tabular}[c]{@{}c@{}}Performance \\ Metric\end{tabular}}} & \multirow{2}{*}{\texttt{DFT1024}} & \multicolumn{2}{c|}{\texttt{ADFT1024\_1}} & \multicolumn{2}{c|}{\texttt{ADFT1024\_2}} & \multicolumn{2}{c}{\texttt{ADFT1024\_3}} \\ \cline{3-8}
\multicolumn{1}{c|}{} &  & Value & Change & Value & Change & Value & Change \\ \hline
Area, $A$ (mm$^2$) & 18.2 & 9.4 & 48.3$\%\downarrow$ & 14.8 & 18.8$\%\downarrow$ & 12.8 & 29.5$\%\downarrow$ \\ \hline
\begin{tabular}[c]{@{}l@{}}Critical Path Delay, $T$ \\ (ns)\end{tabular} & 1.9 & 1.8 & 3.1$\%\downarrow$ & 1.9 & - & 1.9 & - \\ \hline
\begin{tabular}[c]{@{}l@{}}Frequency, $F_{\text{max}}$\\ (GHz)\end{tabular} & 0.5 & 0.6 & 3.2$\%\uparrow$ & 0.5 & - & 0.5 & - \\ \hline
\begin{tabular}[c]{@{}l@{}}$AT$\\ (mm$^2$ns)\end{tabular} & 34.2 & 17.1 & 49.9$\%\downarrow$ & 27.8 & 18.8$\%\downarrow$ & 24.1 & 29.5$\%\downarrow$ \\ \hline
\begin{tabular}[c]{@{}l@{}}$AT^2$\\ (mm$^2$ns$^2$)\end{tabular} & 64.3 & 31.2 & 51.5$\%\downarrow$ & 52.3 & 18.8$\%\downarrow$ & 45.4 & 29.5$\%\downarrow$ \\ \hline
\begin{tabular}[c]{@{}l@{}}Dynamic Power, $D_p$\\ (mW/GHz)\end{tabular} & 44.8 & 15.2 & 66.0$\%\downarrow$ & 36.7 & 18.0$\%\downarrow$ & 23.3 & 48.0$\%\downarrow$ \\ \hline
\end{tabular}
\end{table*}

\subsection{Analysis of the Results}
The results in Table~\ref{comp_table}
shows that the 32-point ADFT core demands
considerably less hardware resources
than the 32-point exact DFT core.
On the other hand,
the
implementation of the transpose buffer
with twiddle factor multiplication adds a
fixed hardware complexity to the system for
both the DFT and the approximate architectures.
As a result, the
transpose buffer
causes the highest area
consumption and a relatively high power consumption
in comparison to that of 32-point ADFT cores.
Thus, it
becomes the dominant factor
in hardware complexity for the designs of the three 1024-point approximate
transforms, as shown
in Table~\ref{algo_table}.

The core~\texttt{ADFT1024\_1} gives the best hardware utilization,
whereas~\texttt{ADFT1024\_2} gives the worst as can be seen in
Table~\ref{algo_table}.
Algorithm~3 gives the best error performance,
i.e., provides the most accurate approximation.
Moreover, the hardware resource consumption of its
physical realization~\texttt{ADFT1024\_3} is also
close to that of~\texttt{ADFT1024\_2}.
The error performance of Algorithm~2 does not differ much
from that of Algorithm~1, which also provides
a hardware realization~\texttt{ADFT1024\_1} with the
lowest resource consumption.
Therefore,
we recommend
either Algorithm~1 or Algorithm~3 (i.e., its hardware realizations~\texttt{ADFT1024\_1} and~\texttt{ADFT1024\_3})
as the best designs.
\subsection{ADFT-based 1024-Beam Digital Beamformers}
\begin{figure*}
\centering
\psfrag{30d}[][][0.7]{30$^\circ$}
\psfrag{60d}[][][0.7]{60$^\circ$}
\psfrag{90d}[][][0.7]{90$^\circ$}
\psfrag{330d}[][][0.7]{330$^\circ$}
\psfrag{300d}[][][0.7]{300$^\circ$}
\psfrag{270d}[][][0.7]{270$^\circ$}
\psfrag{0d}[][][0.7]{0$^\circ$}
\psfrag{BEAM128}[][][0.7]{Beam 128}
\psfrag{BEAM127}[][][0.7]{Beam 127}
\psfrag{BEAM256}[][][0.7]{Beam 256}
\psfrag{BEAM255}[][][0.7]{Beam 255}
\psfrag{BEAM510}[][][0.7]{Beam 510}
\psfrag{BEAM512}[][][0.7]{Beam 512}
\psfrag{BEAM511}[][][0.7]{Beam 511}
\psfrag{BEAM507}[][][0.7]{Beam 507}
\psfrag{BEAM508}[][][0.7]{Beam 508}
\psfrag{BEAM518}[][][0.7]{Beam 518}
\psfrag{BEAM519}[][][0.7]{Beam 519}
\psfrag{0.002}[][][0.7]{0.002}
\psfrag{0.004}[][][0.7]{0.004}
\psfrag{0.006}[][][0.7]{0.006}
\psfrag{0.2}[][][0.7]{0.2}
\psfrag{0.4}[][][0.7]{0.4}
\psfrag{0.6}[][][0.7]{0.6}
\psfrag{0.8}[][][0.7]{0.8}
\psfrag{1.0}[][][0.7]{1.0}
\psfrag{-4.8}[][][0.7]{-4.8}
\psfrag{-1.8}[][][0.7]{-1.8}
\psfrag{0}[][][0.7]{0}
\psfrag{-37.0}[][][0.7]{-37.0}
\psfrag{-30.0}[][][0.7]{-30.0}
\psfrag{-22.2}[][][0.7]{-22.2}
\subfigure[DFT]{\label{fig:BEAMS_POLAR_DFT1}\epsfig{file=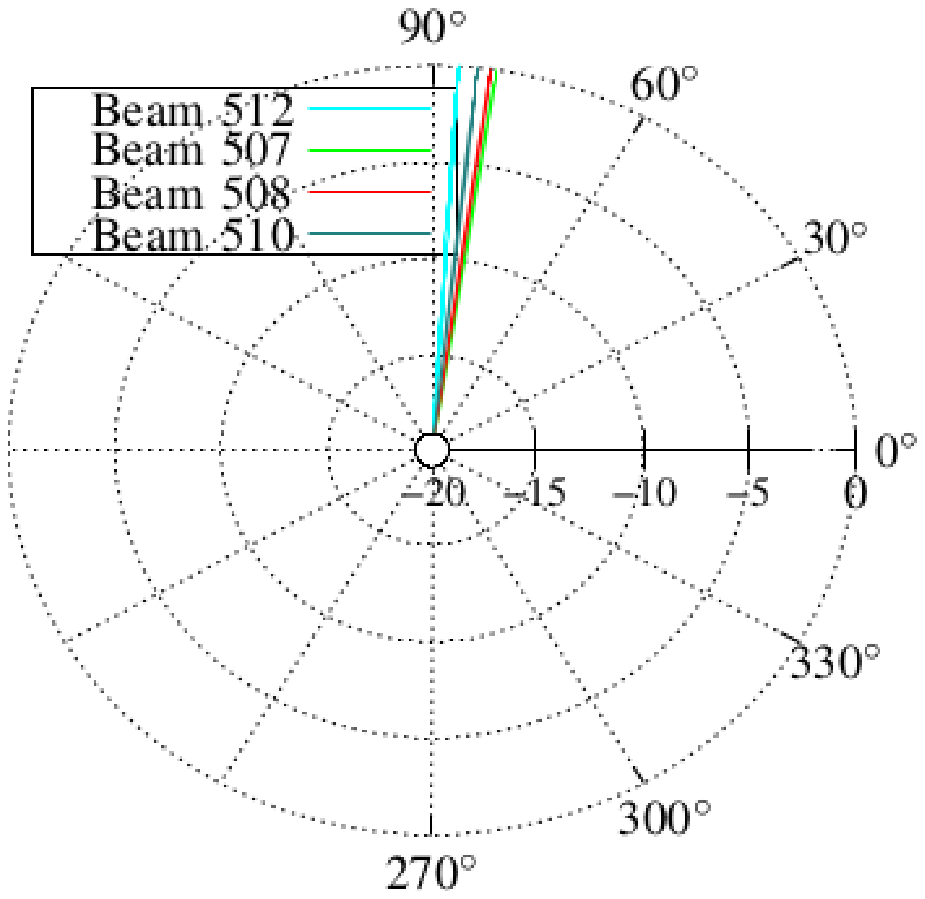, scale=0.5}}
\subfigure[Algorithm~1]{\label{fig:BEAMS_POLAR_ADFT1}\epsfig{file=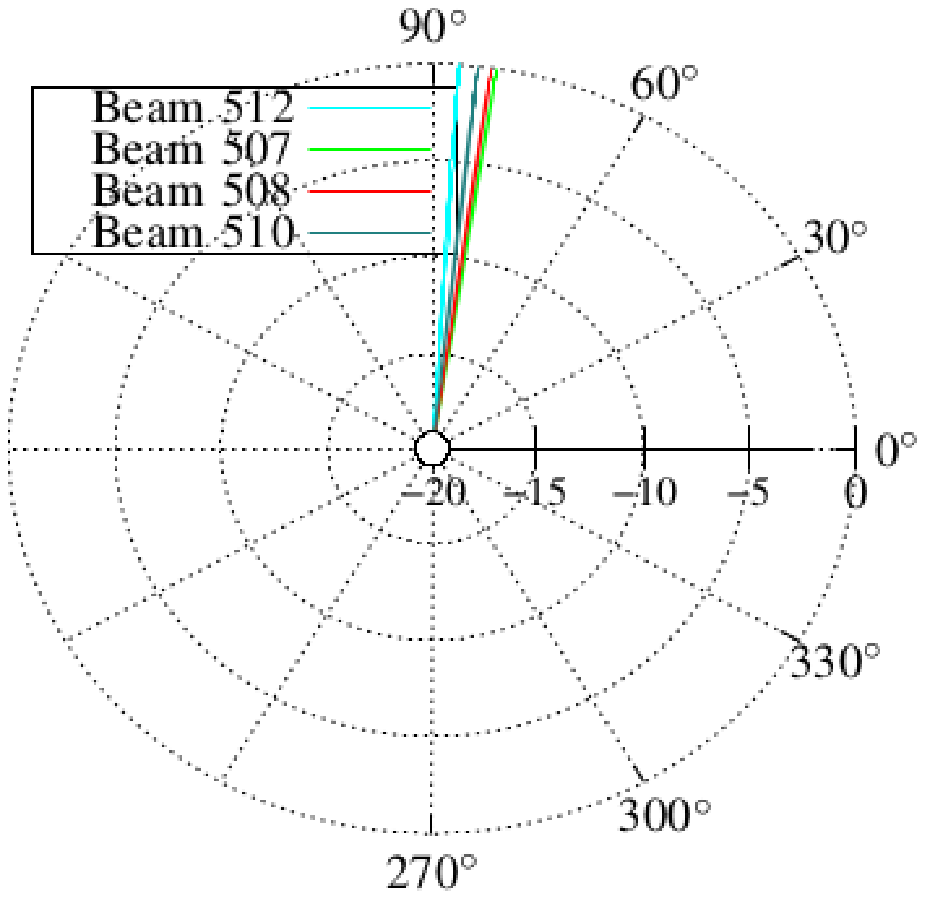, scale=0.5}}
\subfigure[Error for Algorithm~1]{\label{fig:BEAMS_POLAR_EEE1}\epsfig{file=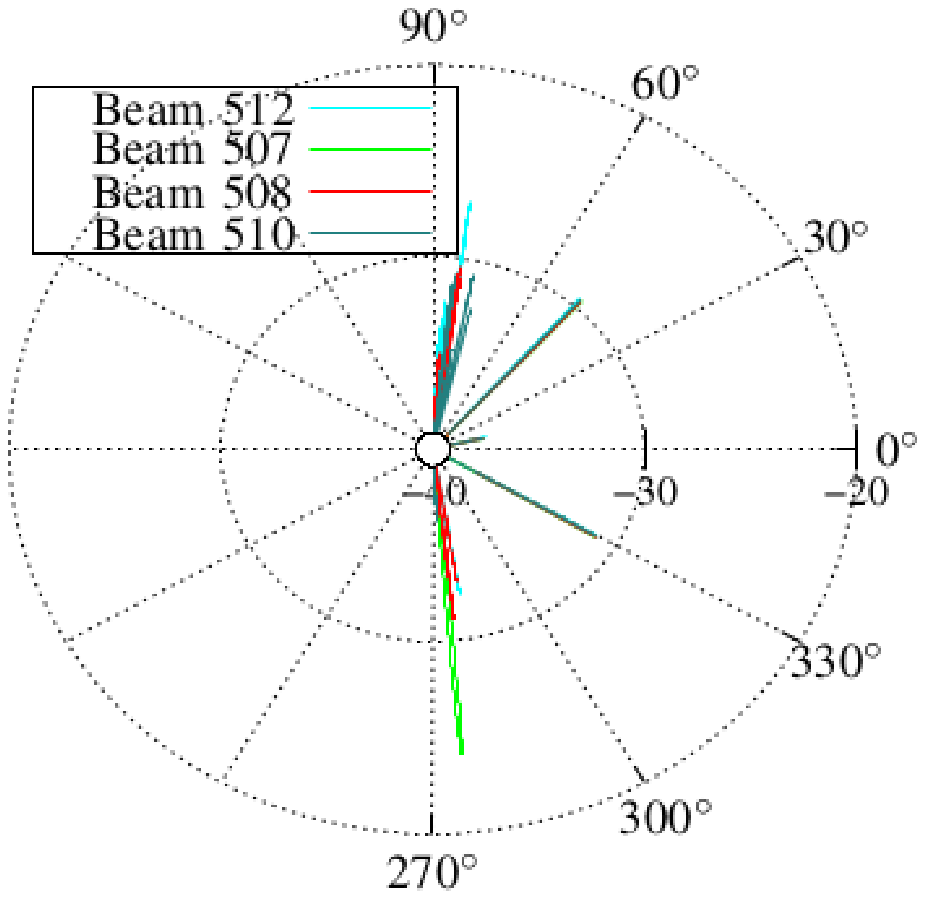, scale=0.5}}
\\
\subfigure[DFT]{\label{fig:BEAMS_POLAR_DFT2}\epsfig{file=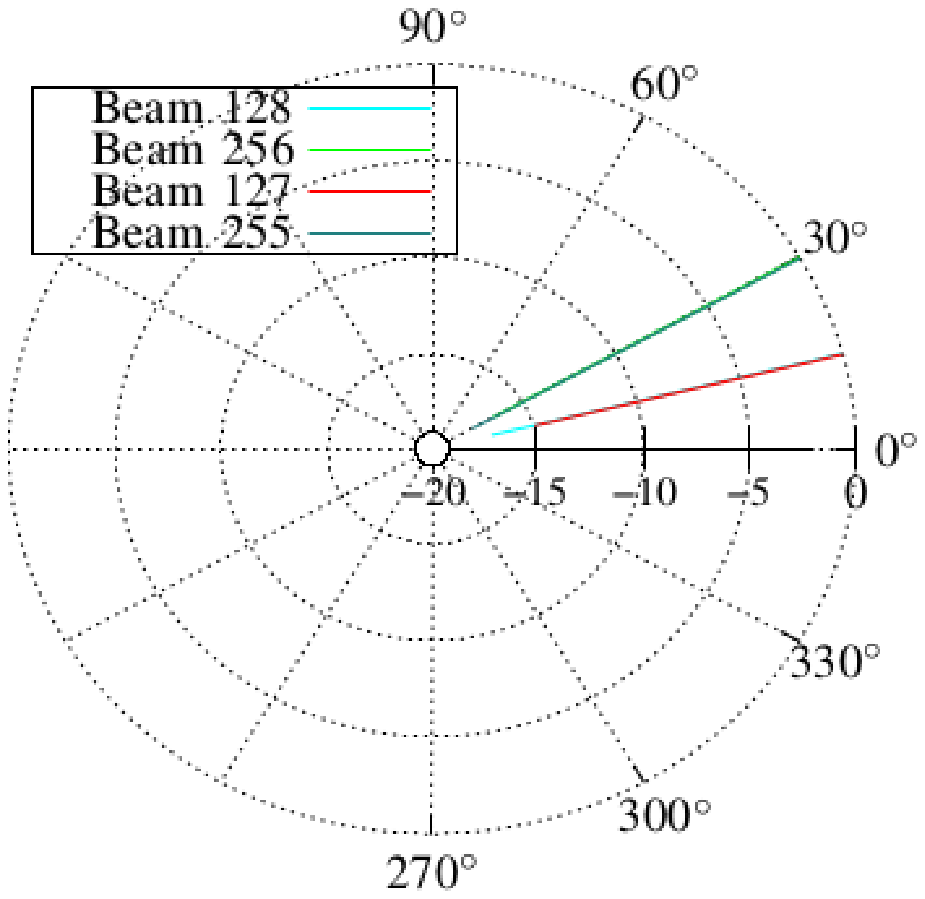, scale=0.5}}
\subfigure[Algorithm~2]{\label{fig:BEAMS_POLAR_ADFT2}\epsfig{file=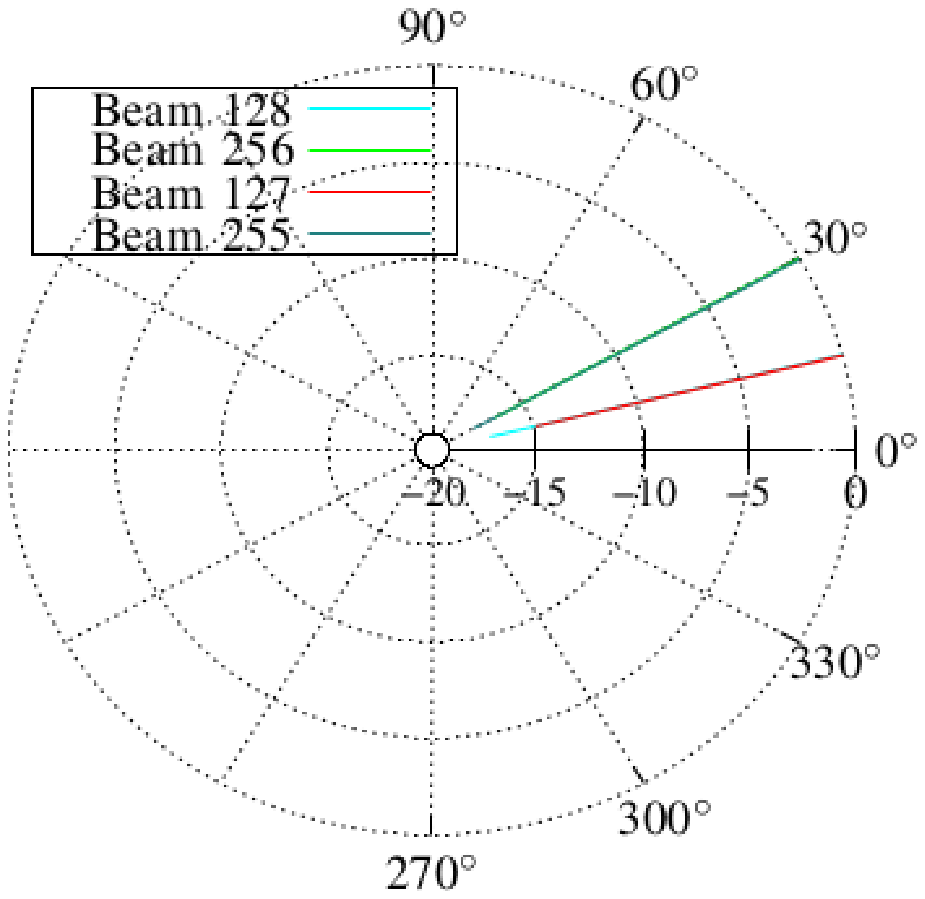, scale=0.5}}
\subfigure[Error for Algorithm~2]{\label{fig:BEAMS_POLAR_ERR2}\epsfig{file=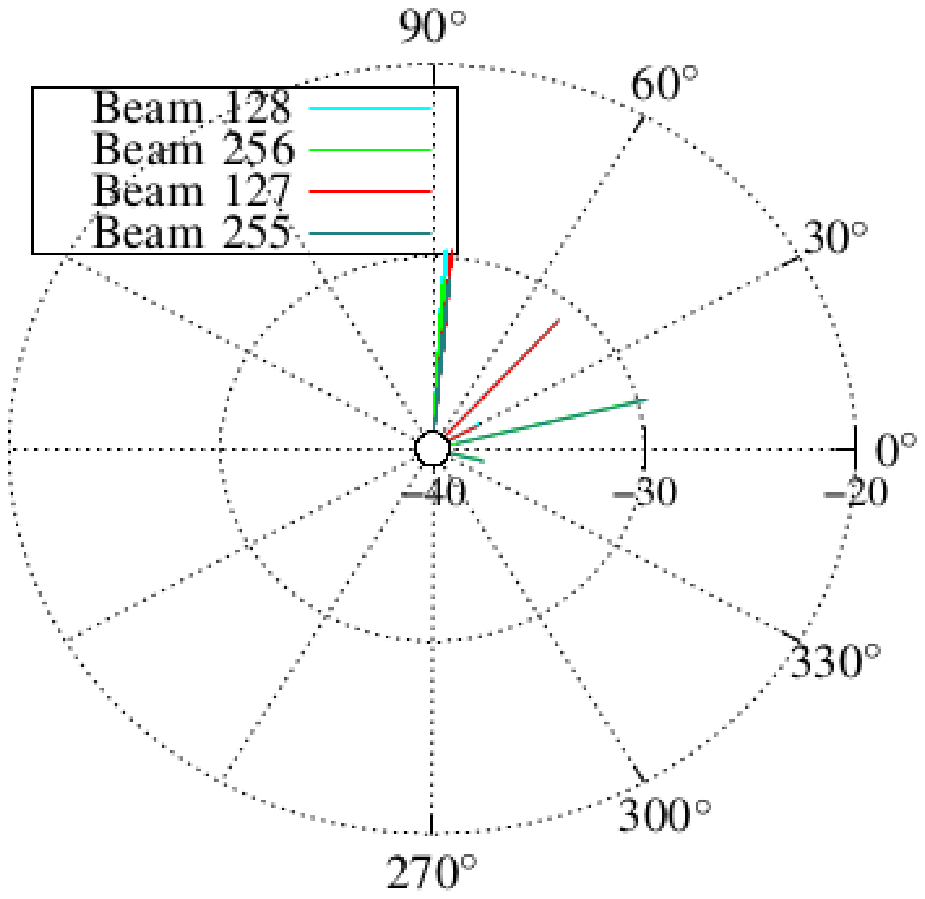, scale=0.5}}
\\
\subfigure[DFT]{\label{fig:BEAMS_POLAR_DFT3}\epsfig{file=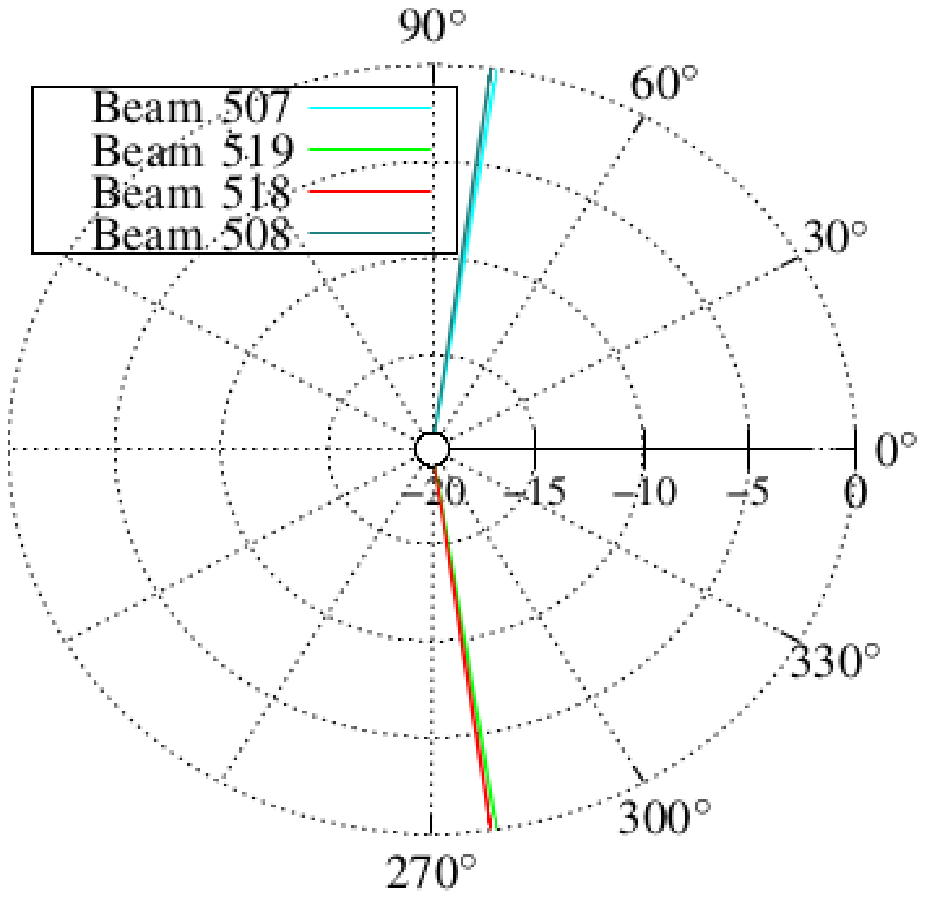, scale=0.5}}
\subfigure[Algorithm~3]{\label{fig:BEAMS_POLAR_ADFT3}\epsfig{file=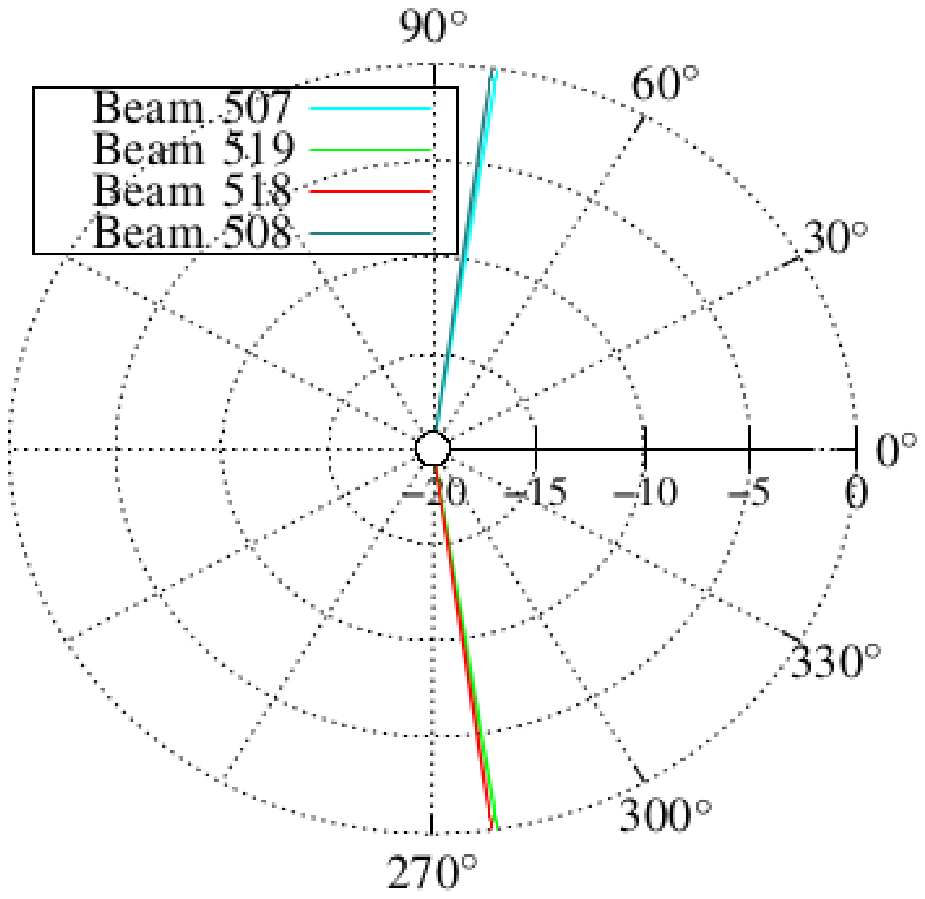, scale=0.5}}
\subfigure[Error for Algorithm~3]{\label{fig:BEAMS_POLAR_ERR3}\epsfig{file=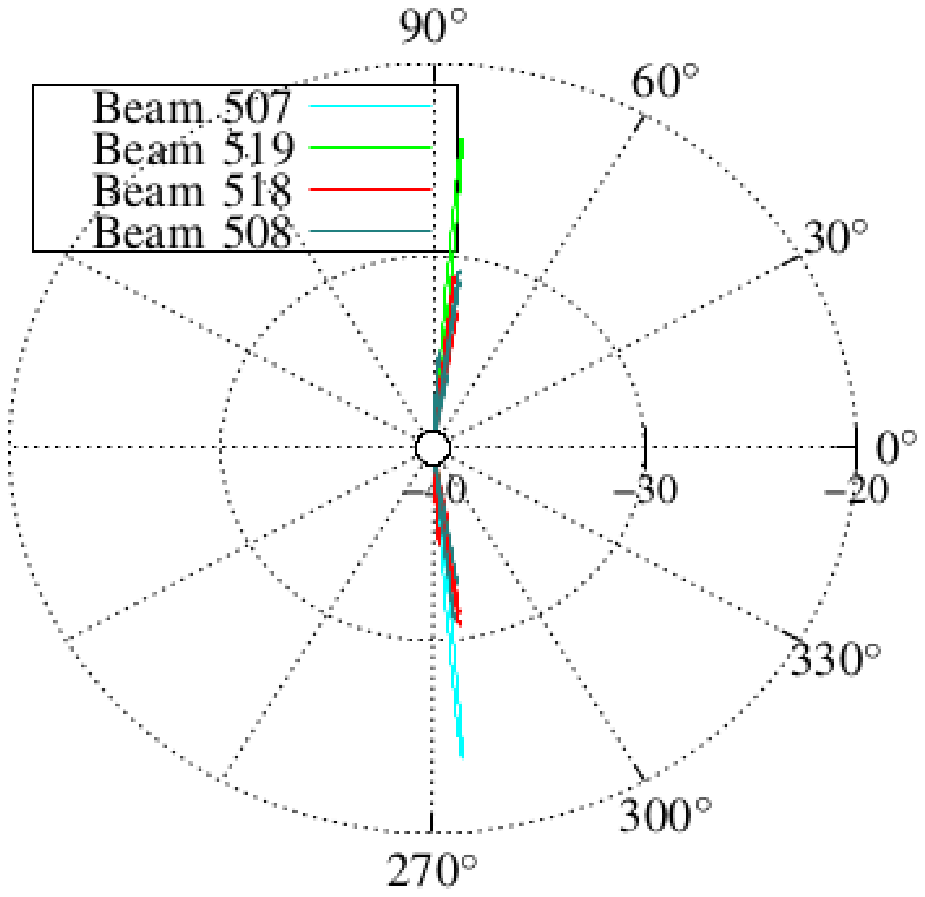, scale=0.5}}

\caption{The four worst bins for multi-beam beamforming: (a) exact DFT response, (b) ADFT response, and (c) error for algorithm 1; (d) exact DFT response, (e) ADFT response, and (f) error for algorithm 2; (g) exact DFT response, (h) ADFT response, and (i) error for algorithm 3. }
\label{fig:beams_polar}
\end{figure*}

 In the proposed system, each ADFT bin corresponds to a unique direction in space. Ideally these bins should be identical to the spatial DFT bins, but their magnitude could deviate because of the approximation. The four worst bins for each of the three algorithms are shown in Fig.~\ref{fig:beams_polar}. The resulting errors are small enough to be acceptable for the low SNR scenarios seen in practical wireless systems.

\begin{figure*}
\centering
\psfrag{30d}[][][0.7]{30$^\circ$}
\psfrag{60d}[][][0.7]{60$^\circ$}
\psfrag{90d}[][][0.7]{90$^\circ$}
\psfrag{330d}[][][0.7]{330$^\circ$}
\psfrag{300d}[][][0.7]{300$^\circ$}
\psfrag{270d}[][][0.7]{270$^\circ$}
\psfrag{0d}[][][0.7]{0$^\circ$}
\psfrag{BEAM128}[][][0.7]{Beam 128}
\psfrag{BEAM127}[][][0.7]{Beam 127}
\psfrag{BEAM256}[][][0.7]{Beam 256}
\psfrag{BEAM255}[][][0.7]{Beam 255}
\psfrag{BEAM510}[][][0.7]{Beam 510}
\psfrag{BEAM512}[][][0.7]{Beam 512}
\psfrag{BEAM511}[][][0.7]{Beam 511}
\psfrag{BEAM507}[][][0.7]{Beam 507}
\psfrag{BEAM508}[][][0.7]{Beam 508}
\psfrag{BEAM518}[][][0.7]{Beam 518}
\psfrag{BEAM519}[][][0.7]{Beam 519}
\psfrag{0.002}[][][0.7]{0.002}
\psfrag{0.004}[][][0.7]{0.004}
\psfrag{0.006}[][][0.7]{0.006}
\psfrag{0.2}[][][0.7]{0.2}
\psfrag{0.4}[][][0.7]{0.4}
\psfrag{0.6}[][][0.7]{0.6}
\psfrag{0.8}[][][0.7]{0.8}
\psfrag{1.0}[][][0.7]{1.0}
\psfrag{-4.8}[][][0.7]{-4.8}
\psfrag{-1.8}[][][0.7]{-1.8}
\psfrag{0}[][][0.7]{0}
\psfrag{-37.0}[][][0.7]{-37.0}
\psfrag{-30.0}[][][0.7]{-30.0}
\psfrag{-22.2}[][][0.7]{-22.2}

\subfigure[DFT]{\label{fig:LINEAR_BEAMS_POLAR_DFT1}\epsfig{file=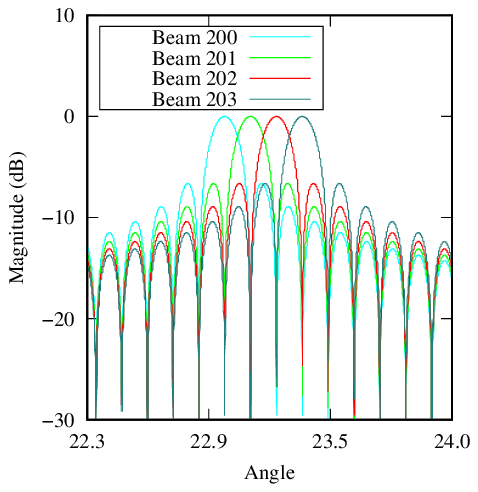, scale=1.0}}
\subfigure[Algorithm~1]{\label{fig:LINEAR_BEAMS_POLAR_ADFT1}\epsfig{file=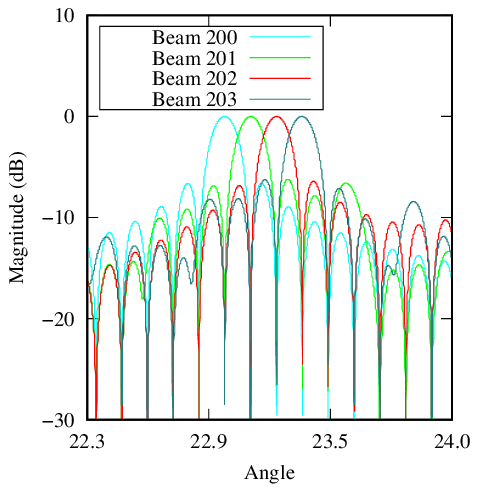, scale=1.0}}
\subfigure[Error for Algorithm~1]{\label{fig:LINEAR_BEAMS_POLAR_ERR1}\epsfig{file=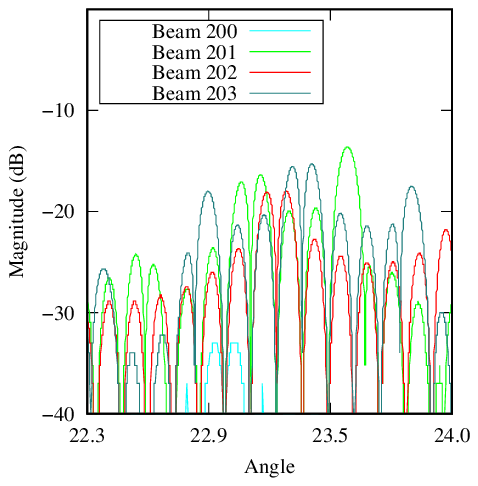, scale=1.0}}
\\
\subfigure[Algorithm~2]{\label{fig:LINEAR_BEAMS_POLAR_ADFT2}\epsfig{file=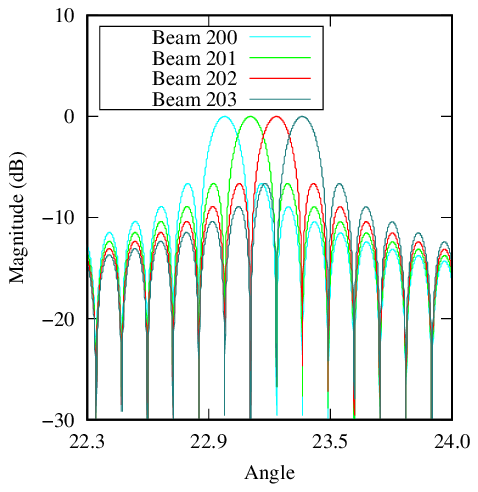, scale=1.0}}
\subfigure[Error for Algorithm~2]{\label{fig:LINEAR_BEAMS_POLAR_ERR2}\epsfig{file=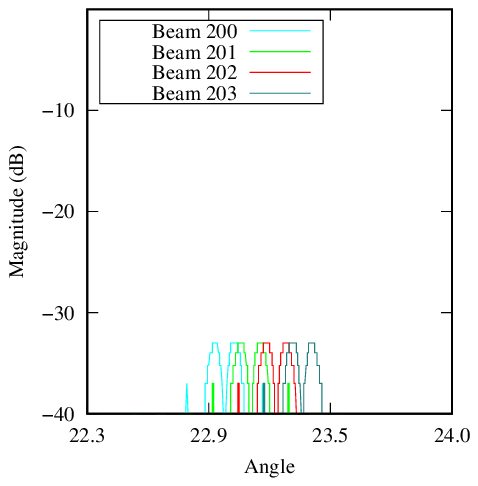, scale=1.0}}
\\
\subfigure[Algorithm~3]{\label{fig:LINEAR_BEAMS_POLAR_ADFT3}\epsfig{file=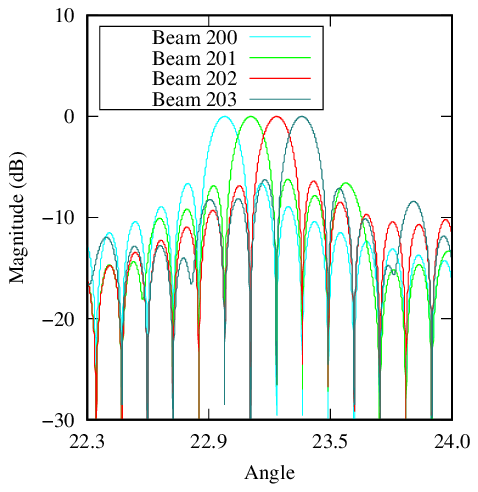, scale=1.0}}
\subfigure[Error for Algorithm~3]{\label{fig:LINEAR_BEAMS_POLAR_ERR3}\epsfig{file=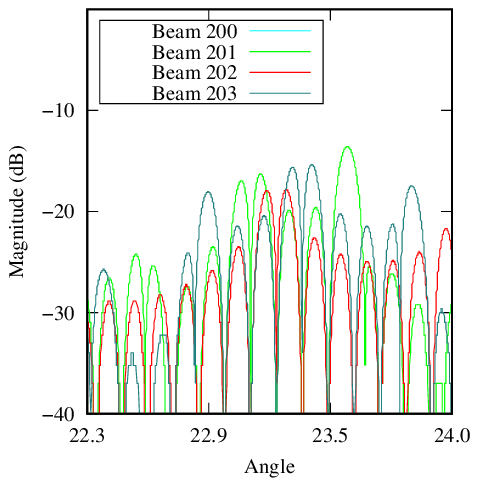, scale=1.0}}

\caption{Linear plot of selected beams $\{200, 201, 202, 203\}$ for exact and approximate transforms: (a) exact DFT response, (b) ADFT response, and (c) error for algorithm 1; (d) ADFT response, and (e) error for algorithm 2; (f) ADFT response, and (g) error for algorithm 3.}
\label{fig:linear_beams_polar}
\end{figure*}

Fig.~\ref{fig:linear_beams_polar} shows detailed plots of 4 consecutive beams (from bins 200-203) for the three proposed algorithms, together with the errors. We chose these four beams arbitrarily to showcase the shapes of the obtained RF beams in sufficient detail.

Note that practical realization of 1024-element ULAs for generating narrow ADFT-based beams in currently-licensed frequency bands (upto the V band) may be challenging due to the large sizes of the resulting apertures. However, due to ongoing research in the sub-THz range~\cite{wb1,wb2,wb3,wb4,wb5,wb6,wb7,wb8,wb9}, the W and G bands will soon be commercially available for both licensed and unlicensed use. At a carrier frequency of 300~GHz, $\lambda/2 = 0.5$~mm and thus the size of a Nyquist-spaced 1024-element ULA would decrease to a reasonable value of 51.2~cm.

\section{Conclusions}
\label{s:conclusion}
FFTs are used for reducing the computational costs
of evaluating the DFT.
Generally, they decrease complexity from~$\mathcal{O} (N^2)$
down to~$\mathcal{O} (N \log N )$.
In this paper,
we show that further savings can be accomplished
by means of approximate methods.
The resulting
1024-point DFT approximations
present a trade-off between performance
and hardware complexity without significant loss in terms of worst-side lobe and SNR.

Our work shows that larger block-length DFT approximations
can be obtained from the smaller-size approximations derived
using previously-described numerical optimization methods.
Our methodology can be directly
applied to any DFT for which the block length is a perfect square.
Since the current DFT approximations in the literature
are restricted to
the sizes $\{8, 16, 32\}$~\cite{Ariyarathna2019,Coutinho2018,Ariyarathna2015, Kulasekera2015, DoraThesis},
approximate algorithms
can be derived for
$N \in \{ 64, 256, 1024 \}$.
In this work,
we focused on the 1024-point case.
Assuming that
a multiplierless DFT approximation of size~$\sqrt{N}$
can always be found,
our derivations suggests that
we can obtain
an $N$-point DFT approximation
that
requires only
$N - 2 \sqrt{N} -1$ multiplications;
effectively
making the complexity
of the resulting $N$-point approximation $\mathcal{O}(N)$.
The proposed algorithms were synthesized to digital VLSI using a 45-nm CMOS library. Synthesis results confirm the expected improvements in layout area and power consumption metrics compared to a conventional 1024-point DFT implementation.

The choice of algorithm depends on the application and its tolerance for computational error in the DFT block. Highly error tolerant applications can greatly benefit from Algorithm 1 which has the lowest complexity. Algorithm 2 or 3 maybe selected when Algorithm 1 does not furnish sufficient performance.

\section*{Acknowledgment}
This work was supported in part by multiple awards from NSF SpecEES and NSF CCSS.
The second author thanks
a careful reviewer and also Mr.~L. Portella
for the identification and correction of a typo
in the matrix factorization shown in Section~2.

%

%


{\small
\singlespacing
\bibliographystyle{siam}
\bibliography{ref}
}

\end{document}